\documentclass[preprint,showpacs,preprintnumbers,amsmath,amssymb]{revtex4}

\usepackage{graphicx}
\usepackage{dcolumn}
\usepackage{bm}

\begin{document}

\title{Spin-transfer in an open ferromagnetic layer: from negative
damping to effective
temperature}
\author{J.-E. Wegrowe, M. C. Ciornei, H.-J. Drouhin}
\affiliation{Laboratoire des Solides Irradi\'es, Ecole
Polytechnique, CNRS-UMR 7642 \& CEA/DSM/DRECAM, 91128 Palaiseau Cedex, France.}

\date{\today}

\begin{abstract}

Spin-transfer is a typical spintronics effect that allows a
ferromagnetic layer to be switched by spin-injection.  Most of the
experimental results about spin transfer (quasi-static hysteresis
loops or AC resonance measurements) are described on the basis of the
Landau-Lifshitz-Gilbert equation of the magnetization, in which
additional current-dependent damping factors are added, and can be
positive or negative.  The origin of the damping can be investigated further by
performing stochastic experiments, like one shot relaxation
experiments under spin-injection in the activation regime of the
magnetization.  In this regime, the N\'eel-Brown activation law is
observed which leads to the introduction of a current-dependent
effective temperature.  In order to justify the introduction of these
counterintuitive parameters (effective temperature and negative
damping), a detailed thermokinetic analysis of the different
sub-systems involved is performed.  We propose a thermokinetic
description of the different forms of energy exchanged between the
electric and the ferromagnetic sub-systems at a Normal/Ferromagnetic
junction.

The derivation of the Fokker-Planck equation in the
framework of the thermokinetic theory allows the damping parameters to
be defined from the entropy variation and refined with the Onsager
reciprocity relations and symmetry properties of the magnetic system.
The contribution of the spin-polarized current is introduced as an
external
source term in the conservation laws of the ferromagnetic layer.  Due
to the relaxation time separation, this contribution can be reduced to
an effective damping.  The flux of energy transferred between the
ferromagnet and the spin-polarized current can be positive or
negative, depending on spin accumulation configuration.  The
effective temperature is deduced in the activation (stationary)
regime, providing that the relaxation time that couples the
magnetization to the spin-polarized current is shorter than the relaxation
to the lattice.

\end{abstract}

\pacs{72.25.Hg, 75.47.De, 75.40.Gb \hfill}

\maketitle

In the context of spintronics, the electrical resistance of magnetic
nanostructures are tuned with the magnetization states.  Giant
magnetoresistance (GMR), or anisotropic magnetoresistance (AMR) allow
the magnetization states of nano-layers to be measured with great
precision.  Such magnetoresistances are easily scalable reading
processes and are used for magnetic sensors and random access memorie
(MRAM) technology.  The possibility of controlling the magnetic
configuration of a magnetic nanostructure by injecting spins emerged
only in recent studies, opening the way to a readily scalable writing
process for MRAMs application.  This approach is also extended to
thermally assisted switching, in which the heat fluxes are also
exploited in order to help the magnetization reversal.  In order to
control the magnetic configurations and their stabilities (for reading
and writing processes), in such magnetic nanopillars, it is necessary
to understand on one hand the processes responsible for the
magnetization reversal (in the presence of a magnetic field and heat),
and on the other hand, the processes governing spin-dependent
electronic transport at normal/Ferromagnetic interfaces.  Taken
separately, both effects are rather well understood today.  However,
coupling the two processes leads one to consider a large variety of
possible mechanisms, called spin-transfer, that involve an ensemble of
non-equilibrium sub-systems in interaction, with different populations
of electrons and different populations of spins.  The present work
tries to clarify this picture with a phenomenological analysis based
on non-equilibrium thermodynamics of open systems.

Magnetization reversal provoked by spin injection has been observed in
magnetic nanostructures of various morphologies, from spin-valve
multilayers \cite{Albert,Myers,Julie,Sun,APL,Kent,Deac} to nanowires
\cite{EPL,Derek,Marcel} or point contacts
\cite{Tsoi0,Tsoi,PRLStiles,Rippard}, and different types of magnetic
domain walls \cite{BergerDW,JulieDW,Vernier,Klaui,SShape,Luc}.  In order
to describe and interpret these observations, physicists where forced
to add one or two current-dependent terms into the well-known
dynamical equations that describe a ferromagnetic layer coupled to a
heat bath (Fokker-Planck or corresponding Landau-Lifshitz-Gilbert
equations).  However, the question remains open about the
deterministic (e.g. spin-torque) or stochastic (e.g. irreversible)
nature of the terms to be added.

It has been observed that for a time window larger than the nanosecond
time scale, and in the framework of one-shot measurements (i.e.
non-averaged, or irreversible measurements), the magnetization
reversal induced by spin-injection is an activated process, with two
level fluctuations \cite{MSU,SPIE,Fabian,Pufall} or simple
irreversible jumps \cite{Guittienne,SPIE}.  In these experiments,
governed by stochastic fluctuations and noise, the observed effect is
accounted for by a current dependent effective temperature in the
N\'eel-Brown activation law \cite{SPIE}.  In contrast, for
quasi-static measurements (e.g. magnetoresistance measured as a
function of the magnetic field or current with DC systems or lock-in
detection system) and for high frequency measurements, oscillations
and resonances indicate, in the frequency domain, the manifestation of
quasi-ballistic precession effects
\cite{Rippard,Pufall,Kiselev,Covington,Krivorotov}.  In these last
experiments, the stochastic nature of the signal is averaged out, and
the behavior is described in terms of current dependent negative
damping within a generalized Landau-Lifshitz-Gilbert (LLG) equation.
This negative damping formulation is motivated by the pioneering works
of Berger \cite{Berger} and Slonczewski \cite{Sloncz} about the
deterministic spin transfer torque theory.  However, the deterministic
approach cannot directly account for the magnetic relaxation
measurements performed in the activation regime (as discussed in Sec.
IV-A below).  The hypothesis of the Slonczewski's spin-torque (presented
as a current-dependent deterministic term in the microscopic
Landau-Lifshitz-Gilbert equation) is not useful as such in the description
proposed here, i.e. in the context of open systems.

In order to justify the introduction of the counterintuitive
phenomenological parameters (effective temperature and negative
damping), a detailed analysis of the different sub-systems is
performed on the basis of thermokinetic theory
\cite{Prigogine53,Prigogine,Guggenheim,Stuck,DeGroot,Kuiken,Parrott,
Mazur,Gruber,Vilar, Rubi2,PRBThermo,FourChan,MTEPW}.  The first step
(first section below) is to identify the relevant sub-systems of
interest ( pointing out the difference between the
spin-accumulation due to the diffusion of spin-dependent conduction
electrons at an interface, and the magnetization of a ferromagnetic
layer), the coupling between them, and the role of microscopic degree
of freedom that will be reduced to the action of the environment.  In
section two, spin-injection and spin-dependent transport are described
in the framework of the two spin-channel approximation (a conduction
channel that carries spin up and a conduction channel that carries
spin down, defined by the conductivities).  Giant magnetoresistance,
spin-accumulation, and corresponding entropy production, or heat
transfer, are deduced.  Beyond the two spin channel approximation, the
analysis is extended to four channels with the introduction of two
other electronic populations (typically $s$-like for conduction
electrons, and $d$-like for the ferromagnetic order parameter) and the
relaxation between them.  In the same manner as spin-flip scattering
coupled the spin-up and spin-down channels, this relaxation defines a
dissipative coupling between the ferromagnet and the spin-dependent
electric sub-systems.  The third section is devoted to the detailed
description of the ferromagnetic order parameter coupled to a heat
bath (without spin-injection).  Both the rotational Fokker-Planck
equation and the corresponding LLG equation are derived in the
framework of the thermokinetic theory, i.e. with the help of the first
two laws of thermodynamics and the Onsager reciprocity relations only.
The coupling of the ferromagnetic order parameter to the heat bath is
introduced via the chemical potential with a typical Maxwell-Boltzmann
diffusion term including the temperature \cite{Prigogine53,Mazur}.
The N\'eel-Brown law is deduced in the activation regime.

The last section is devoted to the ferromagnetic Brownian motion
activated by spin-injection.  The contribution of the spin-polarized
current is introduced by the $s-d$ like relaxation, as a source term
into the conservation laws of the magnetization.  Explicitly, it is
shown that if $n$ is the density of magnetic moments oriented in a
given direction $\theta, \Phi$ of the unit sphere, and $\vec{J}_{M}$
is the corresponding flux of magnetic moments (this flux is not a
displacement in the usual space), the conservation of $n$ writes:
$\partial n/ \partial t = -div \vec{J}_{M} + \int_{N-F} \, \dot
\Psi(z) dz $, where the divergence is defined on the sphere and
$\dot \Psi$ is the relaxation rate, integrated through the
Normal-Ferromagnetic interfaces.  This equation defines the
irreversible spin-transfer occurring in the ferromagnetic layer,
taken as an open system. The relaxation rate is related to the
spin-accumulation $\Delta \mu$ through an Onsager transport
coefficient $L$, $\dot \Psi = L \Delta \mu $ (where $\Delta \mu$ is
proportional to the current).  $L$ is linked to the relaxation times
through the charge conservation laws (or electric screening
properties).

Due to the large relaxation time separation, the contribution of the
source term can be reduced to the effect of an environment that is
responsible for an {\it effective damping} and {\it effective fluctuations}
(or effective
temperature).  The energy transferred between ferromagnetic
layer and the sub-system defined by the
spin-accumulation conduction electrons can be positive or negative,
depending on the sign of the spin accumulation at the different
interfaces.  The effective
temperature is deduced in the activation (stationary) regime, because
the relaxation time that couples the magnetization to the
spin-polarized current short cuts the relaxation to the lattice.

\section{Thermokinetic approach}

\subsection{Interacting sub-systems}
The general scheme of the thermokinetic approach is described in the
references \cite{Prigogine,Stuck,DeGroot,Kuiken,Mazur}.  The method
consists in defining the state of the system with a set of the
relevant extensive variables, say $\{ s, x_{i} \}$, where $x_{i}$
is, e.g. the densities of particles in the sub-system $i$, or
equivalently, the density of component $i$ of a multicomponent
fluid, and $s$ is the total entropy density.  The conservation
equations should then be written, and the two laws of thermodynamics
applied.  The conservation equation for the component $i$ writes:

\begin{equation}
\frac{\partial n_{i}}{\partial t} = -div(\vec J_{i}) + \Sigma_{j}
\nu_{ij} \dot \Psi_{j}
\label{contZero}
\end{equation}

The divergence of the current $\vec J_{i}$ describes the conservative
part of the process, and the term $\dot \Psi_{j}$ is a source term
that describes the relaxation of $\nu_{ij}$ components $i$ into
the component $j$ ($\nu_{ij} \le 0$), or inversely ($\nu_{ij} \ge 0$)
\cite{RquChim}. It is proportional to the inverse of the relaxation
time  $\dot \Psi_{j} \propto \tau^{-1}$ (see Appendix A). Physically,
the term $\dot \Psi_{j}$ describes the relaxation process that changes
the internal degree of freedom (e.g. spins, electric charges,
internal configuration).  In terms of chemical reactions, $\dot
\Psi_{j}$ is the velocity of the reaction, i.e. the generalized flux
thermodynamically conjugated to the chemical affinity $A_{j}$ (defined
below). The summation over all sub-systems, or all components of the
fluid is that of a conserved variable: $\Sigma_{i} \frac{\partial
n_{i}}{\partial t}
= -div(\Sigma_{i} \vec J_{i})$.  The same holds, of course, for the
energy $E$: $\frac{\partial E}{\partial t} = -div(\vec J_{E})$, where
$J_{E}$ is the flux
of energy.  In contrast, the entropy production of the total system
is not conservative in general, due to the
irreversible processes (in other terms, information is lost).  The
equation for the entropy production of the whole system takes the
canonical form $\frac{\partial s}{\partial t} = -div(\vec J_{s}) + \mathcal I$,
where $J_{s}$ is the flux of entropy, and $\mathcal I$ is the
internal entropy production, or {\it irreversibility}, which is a
consequence of the second law of thermodynamics: $\mathcal I \ge 0$
(assuming $T \ge 0$).  According to the first law of thermodynamics,
the energy $E$, is a state function that is also scalar, extensive
and conserved, so that

\begin{equation}
      \frac{\partial E (s,\{ x_{i} \} )}{\partial t}  = \frac{\partial
E}{\partial s}
    \frac{\partial s}{\partial t} + \Sigma_{i} \frac{\partial
    E}{\partial x_{i}} .
\frac{\partial x_{i}}{\partial t}
\label{Prem}
\end{equation}

where $\partial E / \partial s = T$ is the temperature, $\partial E/
\partial x_{i} \equiv F_{i}$ is the generalized force associated with
the flux $\partial x_{i} / \partial t$.  In the following we will
deal exclusively with the chemical potentials $\mu_{i} = \partial
n_{i} /
\partial E$, unless specified otherwise (i.e. there is no need to
introduce other extensive variables).  The following Gibbs
relation is obtained as a direct consequence of the first law: $T
\frac{\partial s}{\partial t} = \frac{\partial E}{\partial t} - \Sigma_{i}
\frac{\partial n_{i}}{\partial t} \mu_{i}$.  After having inserted
the conservation equations, Eq.  (\ref{contZero}), into Eq.
(\ref{Prem}), the following form
is obtained \cite{RqueEcin} :

\begin{equation}
      T \, \frac{\partial s}{\partial t}  = - div (\vec J_{E}) +
\Sigma_{i} \mu_{i}
\, div(\vec
      J_{i}) - \Sigma_{ij} \mu_{i} \, \nu_{ij} \dot \Psi_{j}
\label{EntropOne}
\end{equation}

Using the development $ div(\mu_{i} \, \vec J_{i}) = \mu_{i} \, div(
\vec J_{i}) + \vec J_{i}. \vec{grad}(\mu_{i})$, Eq.
(\ref{EntropOne}) can be re-written in the canonical form:

\begin{equation}
      \frac{\partial s}{\partial t}  = -div(\vec J_{s}) + \mathcal I
\label{EntropTwo}
\end{equation}

where:
\begin{align}
      \left\{
       \begin{aligned}
        \vec J_{s} =& \frac{1}{T}\,\vec J_{E}
             -\Sigma_{i} \frac{\mu_{i}}{T}\,\vec J_{i} \\
        \mathcal{I}=&\, \vec J_{E} \cdot \vec {grad}
            \left(
              \frac{1}{T}
            \right)
         - \Sigma_{i} \vec J_{i} \cdot \vec{grad}
           \left(
             \frac{\mu_{i}}{T}
           \right) - \frac{1}{T} \Sigma_{ij} \, \mu_{i} \, \nu_{ij}
\dot \Psi_{j}
       \end{aligned}
      \label{SProdZero}
      \right.
\end{align}

where the last term on the right hand side defines the dissipative
coupling between the sub-systems. As will be shown in the last
section, this term is responsible for the irreversible
spin-transfer effect described in this work.
What is unusual in dealing with the second law, is to manipulate an
inequality instead of an equality, and consequently to deal with
sufficient conditions instead of equivalences.  Here, the condition
$\mathcal I \ge 0$ leads to a positive matrix $\{
\mathcal L_{ij}\}_{ij}$ of Onsager-Casimir transport coefficients that are
state functions of the variables $\{s,x_{i}\}$, in order to
build a positive quadratic form.  The condition is fulfilled if the
flux $J_{i}$ and the relaxation velocity $\dot \Psi_{i}$ have the form

\begin{align}
      \left\{
       \begin{aligned}
        \vec J_{i} \, = & - \Sigma_{j} \mathcal L_{ij} \, \vec{grad}(\mu_{j})\\
        \dot \Psi_{i} \, =&  \Sigma_{j} L_{i j} \, A_{j}
       \end{aligned}
      \label{Flux}
      \right.
\end{align}

where 

\begin{equation} 
    A_{j} \equiv - \Sigma_{k} \nu_{ik} \, \mu_{k}
 \end{equation}

is the chemical
affinity of the corresponding reaction $j$ (and we have $A_{j} = -
\partial E/ \partial \Psi_{j} $) \cite{DeDonder}.  Furthermore, due
to the time reversal
symmetry of the microscopic equations, the transport coefficients
follow the Onsager-Casimir reciprocity relations \cite{Onsager0}.
The cross-coefficients that couple the flux $\vec J $ to the
relaxation process $\dot \Psi$ are assumed to be zero, because,
according to the Curie principle, only processes of identical
tensorial nature are coupled.
Inserting Eq. (\ref{Flux}) into the continuity equation Eq. (\ref{contZero}),
we obtain an equation of the time variation of the density $\partial
n_{i}/\partial t$ in terms of derivatives of the chemical
potentials $\mu_{j}$:

\begin{equation}
\frac{\partial n_{i}}{\partial t} =  \Sigma_{j}  \mathcal L_{ij} \,
\nabla^{2} \, \mu_{j}
+ \Sigma_{jk} \nu_{ij} L_{i k} \, A_{k}
\label{contFPE}
\end{equation}

It is then sufficient to know the form of the chemical potential as a
function of the density (for pur fluids : $\mu(n_{i}) = \mu_{0} + kT
\, ln(n_{i}/N)$) in order to derive the corresponding differential
equation, or Fokker-Planck
equation, with diffusion and relaxation terms (see sections II, III and VI below).

What we gain in performing this analysis is to identify clearly the
conservative and dissipative flux (through the internal entropy
production), and to be able to define a dissipative process that
couples the sub-systems beyond the usual deterministic coupling
(electric field, magnetic field, pressure, etc\ldots).  This
dissipative coupling appears with an additional transport
coefficient $L$, defined univocally via the transport equations.  In
the case studied below, the matrix $\mathcal L$ is composed by
the\ss conductivities $\sigma_{i}$ associated to each channel (i.e.
associated to a given electronic population), the thermal
conductivity, or the corresponding Seebeck (thermoelectric power)
and Peltier coefficients \cite{MTEPW,Shi,Gravier} and the
ferromagnetic transport coefficients: gyromagnetic ratio $\Gamma$
and the Gilbert damping coefficient $\eta$.  Beyond, the flux of
entropy or heat allows the {\it spin transfer} to be understood in
an open system in terms of relaxation with a supplementary Onsager
coefficient $L$.  As shown in the last section, this term is
responsible for an effective temperature $T_{eff}$ and effective
(negative) damping $\alpha_{eff}$.

\subsection{The model}

The model is based on the hypothesis that the ferromagnetic order
parameter $\vec M$ is well differentiated from the sub-system composed
by spin-polarized conduction electrons, although both systems exchange
charges, spins, and heat through a relaxation mechanism that will be
described in terms of internal variables \cite{Prigogine53,DeGroot,Mazur}.  As
shown above, the relaxation of an internal variable (or
internal degree of freedom) defines a transport coefficient $L_{sd}$
related to the corresponding relaxation time $\tau_{sd}$ ($L_{sd} \propto
\tau_{s d}^{-1}$, see appendix A for the relation to the relaxation
time).

We hence start with the two sub-systems: the ferromagnet described by
the magnetization $\vec M$ and the two conducting spin-channel system
of the conduction electrons.  Both sub-systems are dynamically coupled
through the relaxation time $\tau_{sd}$.  This relaxation is qualified
as interband relaxation, to be opposed to the intraband spin-flip
relaxation $\tau_{sf}$ introduced in the usual two spin-channel
approximation.  The conducting channels are usually described by the
density $n_{\uparrow}$ of conduction electrons with spin up and the
density $n_{\downarrow}$ of conduction electrons with spin down.  The
intraband coupling (accounted for by $L_{sf}$ or $\tau_{sf}$) is
responsible for the spin-accumulation mechanism at stationary
regime.  For convenience, we redefine the two channels with the
density of spin-polarized electrons $\Delta n = n_{\uparrow} -
n_{\downarrow}$ ("spin conduction channel") and the total density of
electrons $n_{0} = n_{\uparrow} + n_{\downarrow}$.

Furthermore, the conduction channels are contacted to a
power supply (current generator here). Strictly speaking, the
magnetic system is also contacted to the power supply, e.g. through
the electron of $d$ character \cite{Stearn}. The conduction electrons
are thermalized each-other through a well-known mechanism of elastic
scattering $\tau_{e}$ (that defines the conduction electron
reservoir), at the femto-second time scales (or below), and are
also contacted to the lattice through the Fermi-Dirac distribution,
and inelastic scattering $\tau_{ph}$. On the other hand, the
ferromagnetic order parameter is contacted to the
lattice with a well-known relaxation time $\tau_{0}$ that is measured in
ferromagnetic resonance (FMR) experiments, and is typically of the order of the
nanosecond (or few hundreds of picoseconds). This description leads to
the model depicted in Fig 1(b).

\begin{figure}
        \begin{center}
         \begin{tabular}{c}
         \includegraphics[height=8cm]{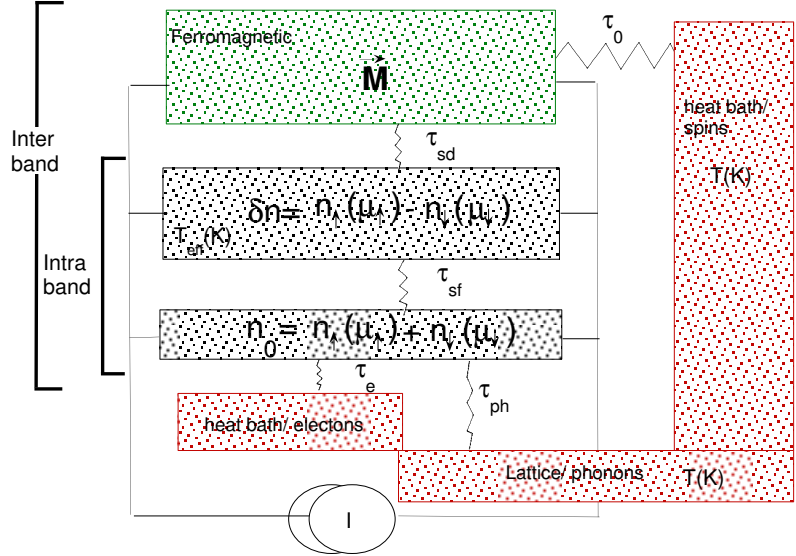}
         \end{tabular}
         \end{center}
         \caption
{ \label{model} Thermokinetic picture of irreversible spin-transfer.
Ferromagnetic system (with magnetization M), and electric system with
spin accumulation density $\Delta n$ and electronic density at the Fermi level
$n_{0}$.  The chemical potential $\mu$ is defined for each spin channel.  The
three sub-systems are coupled together through the relaxation times
$\tau_{sd}$ (interband $s-d$ like relaxation) and
$\tau_{sf}$.(intraband spin-flip
relaxation).  The sub-systems are also coupled to the current
generator $I$, and to the heat reservoirs, through the corresponding
well known relaxation times $\tau_{0}$ (N\'eel-Brown waiting time),
$\tau_{e}$ and
$\tau_{ph}$: elastic and inelastic electronic relaxation times. }
         \end{figure}

The basic idea developed below lays on the fact that the typical time
scales of the dynamics of the two sub-systems are largely separated.
There is a slow variable, the magnetization, and fast variables, the
degree of freedom related to the spin of the conduction electrons.  It
is then possible to reduce the action of the fast variable to the role
of an environment with regard to the magnetization, like for spin-bath
relaxation.  The effect of the coupling to the spin-dependent
electronic sub-system will then be reduced to specific damping and
fluctuation terms added to the usual stochastic equations for the
magnetization.  This will be our line of reasoning followed in the
last section, after describing the two sub-systems.

\section{spin-dependent transport}

In order to explain the high resistance and the high thermoelectric
power observed in transition metals, Mott introduced the concept of
spin-polarized current and suggested that s-d interband scattering
plays an essential role in the conduction properties \cite{Mott}.
This approach in terms of two conduction bands \cite{Stearn},
explained the existence of a spin-polarized current in the 3d
ferromagnetic materials \cite{TwoChan}, and was used for the
description of anisotropic magnetoresistance (AMR)
\cite{Potter0,Potter}, the description of spin-polarizer
\cite{Drouhin}, and thermoelectric power \cite{Handbook}.  With the
discovery of giant magnetoresistance (GMR) \cite{GMR} and related
effects \cite{Awschalom} (like domain wall scattering
\cite{Viret,LevyDW,Ulrich,DWS} discussed below ), the development of
spintronics focused the discussion on spin-flip scattering occurring
between spin-polarized conducting channels
\cite{Gijs,Buttler,Levy0,Zutic,Schmidt,Marrows}.  The two-channel
model, which describes the conduction electrons with majority and
minority spins, is applied with great efficiency to GMR and spin
injection effects
\cite{Johnson,Wyder,Valet,Levy,FertDuvail,Heide,PRBThermo}, including
metal/semiconductor \cite{Molenkamp} and metal/superconductor
interfaces \cite{JedemaSupra}.  In this context, it is sufficient to
describe the diffusion process in terms of spin-flip scattering
without the need to invoke interband s-d scattering.

It is convenient to generalize the two spin channel approach to any
relevant transport channels, i.e. to any distinguishable electron
populations $\alpha$ and $\gamma$ (defined by an internal degree of
freedom).  The local
out-of-equilibrium state near the junction is then described by a
non-vanishing chemical-potential difference between these two
populations: $\Delta \mu_{\alpha \gamma} = \mu_{\alpha}-\mu_{\gamma}
\neq 0$.  In other words, assuming that the presence of a junction
induces a deviation from the local equilibrium, the $\alpha$ and
$\gamma$ populations can be {\it defined by the $\alpha \rightarrow
\gamma$ relaxation mechanism} itself, that allows the local
equilibrium to be recovered in the bulk material ($lim_{z \rightarrow
\pm \infty}\Delta \mu(z) = 0$) \cite{PRBThermo}.  Such considerations
have been presented in some important spintronics studies on the basis
of microscopic calculations
\cite{FertDuvail,Heide,Mott,Potter0,Potter,Gijs,Levy0,Suzuki,Tsymbal,Baxter}.
      The thermokinetic approach \cite{cond-mat} allows us to deal with
interband relaxation on an equal footing with spin-flip relaxation,
with the help of the transport coefficients only.  For this purpose,
the two spin-channel model is generalized, with the introduction of
the corresponding transport coefficients: the conductivities
$\sigma_{\alpha}$ and $\sigma_{\gamma}$ of each channel define the
total conductivity $\sigma_{t}=\sigma_{\alpha} + \sigma_{\gamma}$ and
the conductivity asymmetry $\beta = (\sigma_{\alpha}- \sigma_{\gamma})
/\sigma_{t}$; the relaxation between both channels is described by the
parameter $L$ (or equivalently, the relevant relaxation times
$\tau_{\gamma \leftrightarrow \alpha}$).

\subsection{The generalized two channel model}

\begin{figure}
        \begin{center}
         \begin{tabular}{c}
         \includegraphics[height=5cm]{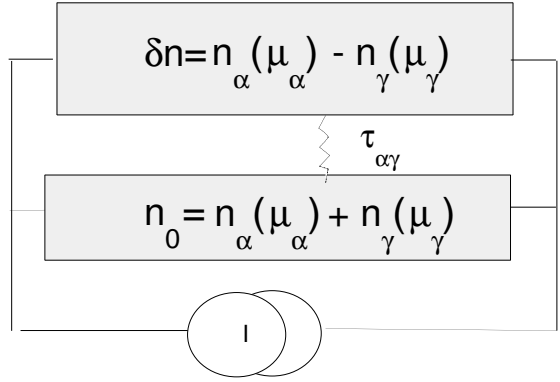}
         \end{tabular}
         \end{center}
         \caption
{ \label{sf} Two channel model, including relaxation
that couples the two electronic populations.}
         \end{figure}

In the framework of the two conducting-channel model, which includes
relaxation from one channel to the other, it easy to follow step by
step the method described in the first section. The conservation laws write
(assuming a 1D space variable $z$):

\begin{equation}
\left\{
          \begin{aligned}
\frac{\partial n_{\alpha}}{\partial t}\, = \,-\frac{\partial
J_{\alpha}}{\partial
z} - \, \dot{\Psi}_{\alpha \gamma} \\
\frac{\partial n_{\gamma}}{\partial t}\, = \,-\frac{\partial
J_{\gamma}}{\partial
z}
+ \, \dot{\Psi}_{\alpha \gamma} \\
\end{aligned}
\label{con0}
\right.
\end{equation}

where $n_{\alpha}$ and $n_{\gamma}$ are  the densities of particles
in the channels $\{\alpha, \gamma \}$.

The entropy variation writes:
\begin{equation}
T \mathcal I = -J_{\alpha} \frac{\partial \mu_{\alpha}}{\partial
z} - J_{\gamma} \frac{\partial \mu_{\gamma}}{\partial z} - \dot
\Psi_{\alpha \gamma} (\mu_{\alpha} - \mu_{\gamma})
\label{entropyTC}
\end{equation}

the application of the second law of thermodynamics leads to
introduce the Onsager coefficients $\sigma_{\alpha} \ge 0$,
$\sigma_{\gamma} \ge 0$ , and $L  \ge 0$
\cite{PRBThermo,cond-mat}, such that:

\begin{equation}
\left\{
\begin{aligned}
J_{\alpha} &= -\frac{\sigma_{\alpha }}{e} \frac{\partial
\mu_{\alpha}}{\partial z}\\
J_{\gamma}& = -\frac{\sigma_{\gamma}}{e} \frac{\partial
\mu_{\gamma}}{\partial z}\\
\dot{\Psi}_{\alpha \gamma}& = L \left ( \mu_{\alpha}-\mu_{\gamma}
\right )
\end{aligned}
\right. \label{Onsager0}
\end{equation}

where $\dot{\Psi}_{\alpha \gamma}$ describes the relaxation from the
channel $\alpha$ to the other channel $\gamma$ in terms of velocity
of the reaction $\alpha \rightarrow \gamma$.  It is not necessary,
in what follows, to distinguish between the electric part and the
pure chemical part of the electro-chemical potentials (see
\cite{FertLee} ). The effects of the electric charge distribution are
described in Appendix A, with the introduction of the screening
length $l$ and the relation to the relaxation times.  As shown in
Appendix A, the Onsager coefficient $L$ is inversely proportional to
the electronic relaxation times $\tau_{\alpha \leftrightarrow
\gamma}$ :

\begin{equation}
L \propto \left ( \frac{g}{\tau_{\alpha \rightarrow \gamma}} +
\frac{f}{\tau_{\gamma \rightarrow \alpha}}\right ) \label{RelaxTau}
          \end{equation}

where $f$ and $g$ are two functions close to unity, and related to the
electric charge distributions (see Appendix A).  Note that due to our
definition of $\mu_{\alpha}$ and $\mu_{\gamma}$, there is no direct
coupling between the two channels : there is no transport coefficients
that couples the two first equations in Eq.  (\ref{Onsager0}).  This
is a consequence of the definition of the electronic populations,
through the relaxation process itself (the populations are stable if
$\dot \Psi= 0$).  Indeed, the out-of-equilibrium configuration at the
interface is quantified by the chemical affinity $\Delta \mu =
\mu_{\alpha} - \mu_{\gamma}$, i.e. the chemical potential difference
of the reaction.

The total current $J_{t}$ is
constant:

        \begin{equation}
J_{t} = J_{\alpha} + J_{\gamma} = -\frac{1}{e} \frac{\partial
}{\partial z} \left (\sigma_{\alpha } \mu_{\alpha}+ \sigma_{\gamma }
\mu_{\gamma } \right )
\end{equation}

However, it is not possible to measure separately the different
conduction channels, since any realistic electric contact short-cuts
the two channels.  What is measured is necessarily the usual Ohm's
law, $J_{t}= -\sigma_{t} \frac{\partial \Phi}{\partial z}$, that
imposes the reference electric potential $\Phi$ to be introduced,
together with the total conductivity $\sigma_{t}=\sigma_{\alpha}+
\sigma_{\gamma}$ \cite{Constantes}. The potential $\Phi$ is hence:

\begin{equation}
e \Phi= \frac{1}{\sigma_{t}}( \sigma_{\alpha} \mu_{\alpha} +
\sigma_{\gamma} \mu_{\gamma})
\end{equation}

\begin{figure}
        \begin{center}
         \begin{tabular}{c}
         \includegraphics[height=6cm]{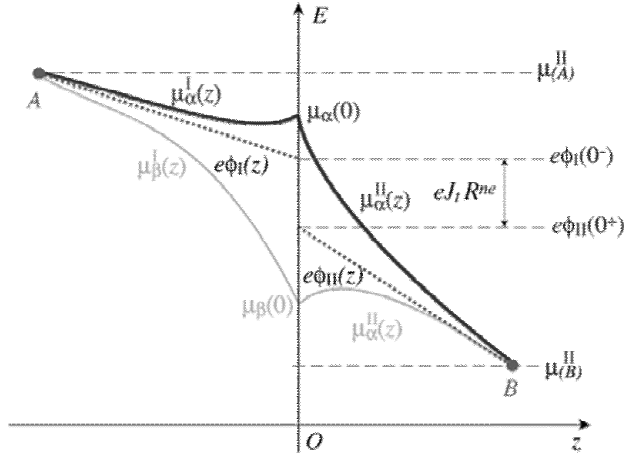}
         \end{tabular}
         \end{center}
         \caption[Chemical potential profile in the $\alpha$ and
$\gamma$ channels]
{ \label{chimique} Junction between to layers $I$ and $II$. Chemical
potential profile over the interval $\left[ A,B\right] $ in the $
\alpha $ and $\gamma $ channels.  The A and B points verify $\mu
_{\alpha }(A)=\mu _{\gamma }(A)$ and $\mu _{\alpha }(B)=\mu _{\gamma
}(B)$.  The two straight lines represent the $\Phi $ variation in
each region ($\Phi_{I} $, $\Phi_{II}$).  It can be directly seen
that the out-of-equilibrium resistance $R^{ne}$ is determined by the
$\Phi $ discontinuity at the interface.}
         \end{figure}

Let us assume that the two channels collapse to a unique conduction
channel for a specific configuration, the reference, which is a
local equilibrium situation: $\Delta \mu_{eq}=0$.  The
out-of-equilibrium contribution to the resistance, $R^{ne}$, is
calculated through the relation:

\begin{equation}
J_{t}e \, R^{ne}  = \int_{A}^{B}  \frac{\partial }{ \partial z}
(\mu_{\alpha} - e \Phi(z))dz = \int_{A}^{B}  \frac{\partial }{
\partial z} (\mu_{\gamma} - e \Phi(z))dz \label{Integ}
\end{equation}
so that
\begin{equation}
        R^{ne}= -\frac{1}{J_{t}e} \int_{A}^{B}
\frac{\sigma_{\alpha} - \sigma_{\gamma}}{2 \sigma_{t}}
\frac{\partial \Delta \mu}{ \partial z}dz \label{Res}
\end{equation}
where the measurement points $A$ and $B$ are located far enough from
the interface (inside the bulk) so that $\Delta \mu (A)=\Delta \mu
(B) =0$ (see Fig. \ref{chimique}).  The integral in Eqs.
(\ref{Integ}) is performed over the regular part of the function
only ($\Phi$ and $\sigma_{i}$ are discontinuous) \cite{Integral}.
Eq.  {\ref{Res}} allows the out-of-equilibrium resistance at a
simple junction between two layers (composed by the layers $I$ and
$II$) to be easily calculated.  If the junction is set at $z=0$ and
the conductivities are respectively $\sigma_{i}^{I}$ and
$\sigma_{i}^{II}$ ($i=\{\alpha, \gamma \}$), we have:

\begin{equation}
J_{T}e \, R^{ne} = \int_{A}^{0} \frac{\sigma^{I}_{\alpha} -
\sigma^{I}_{\gamma}}{2 \sigma_{t}} \frac{\partial \Delta \mu^{I}}{
\partial z}dz + \int_{0}^{B} \frac{\sigma^{II}_{\alpha} -
\sigma^{II}_{\gamma}}{2 \sigma_{t}} \frac{\partial \Delta \mu^{II}}{
\partial z}dz \label{ResJunct}
\end{equation}
The equilibrium is recovered in the bulk, so that:

\begin{equation}
R^{ne} = \left ( \frac{\sigma^{I}_{\alpha} -
\sigma^{I}_{\gamma}}{\sigma^{I}_{t}} - \frac{\sigma^{II}_{\alpha} -
\sigma^{II}_{\gamma}}{\sigma^{II}_{t}} \right ) \frac{\Delta
\mu(0)}{2 J_{t}e} \label{Result}
\end{equation}

The chemical potential difference $\Delta \mu(z)$, which accounts for
the pumping force opposed to the relaxation $\alpha \rightarrow
\gamma$, is obtained by solving the diffusion equation deduced from
Eqs.  (\ref{Onsager0}) and (\ref{con0}), and assuming a stationary
regime for each channels, $\frac{\partial n_{\alpha}}{\partial t}\, =
\frac{\partial
n_{\gamma}}{\partial t} = 0 $
\cite{Johnson,Wyder,Valet,Levy,PRBThermo}:

\begin{equation}
\frac{\partial^{2}\Delta \mu(z)}{\partial z^{2}}= \frac{\Delta
\mu(z)}{l_{diff}^{2}} \label{DiffEq}
\end{equation}
where
\begin{equation}
l_{diff}^{-2}= eL(\sigma_{\alpha}^{-1}+\sigma_{\gamma}^{-1})
\label{ldiff}
\end{equation}
is the diffusion length related to the $\alpha \rightarrow \gamma$
relaxation.

At the interface ($z=0$), the continuity of the currents for each
channel writes $J_{\alpha}^{I}(0)=J_{\alpha}^{II}(0)$, were

\begin{equation}
J_{\alpha}(0)=-\frac{\sigma_{\alpha} \sigma_{\gamma}}{e \sigma_{t}}
\frac{\partial \Delta \mu}{\partial
z}+\frac{\sigma_{\alpha}}{\sigma_{t}} J_{t}
\label{currentcon}
\end{equation}
which leads to the general relation:

\begin{equation}
        \Delta \mu (0)= \left ( \frac{\sigma^{I}_{\alpha}}{\sigma^{I}_{t}} -
\frac{\sigma^{II}_{\alpha}}{\sigma^{II}_{t}} \right ) \, \left (
\frac{\sigma^{I}_{\alpha} \sigma^{I}_{\gamma}}{\sigma^{I}_{t}
l^{I}_{diff}}+
\frac{\sigma^{II}_{\alpha}\sigma^{II}_{\gamma}}{\sigma^{II}_{t}
l^{II}_{diff}} \right )^{-1}\, \, eJ_{t} \label{DeltaMu0}
\end{equation}

Inserting Eq. (\ref{DeltaMu0}) into Eq.  (\ref{Result}), we obtain
the general expression for the out-of-equilibrium resistance (per
unit area) produced by the $\alpha \rightarrow \gamma$ relaxation
mechanism at a junction:

\begin{equation}
R^{ne} = \left ( \frac{\sigma^{I}_{\alpha} - \sigma^{I}_{\gamma}}{2
\sigma^{I}_{t}} - \frac{\sigma^{II}_{\alpha} -
\sigma^{II}_{\gamma}}{2 \sigma^{II}_{t}} \right )\, \left (
\frac{\sigma^{I}_{\alpha}}{\sigma^{I}_{t}} -
\frac{\sigma^{II}_{\alpha}}{\sigma^{II}_{t}} \right ) \, \left (
\sqrt{\frac{\sigma^{I}_{\alpha}
\sigma^{I}_{\gamma}eL^{I}}{\sigma^{I}_{t}}} +
\sqrt{\frac{\sigma^{II}_{\alpha}
\sigma^{II}_{\gamma}eL^{II}}{\sigma^{II}_{t}}}\right )^{-1}
\label{Rout}
\end{equation}

where we have used the relation :

\begin{equation}
          l_{diff}^{-1}= 2 \sqrt{\frac{eL}{\sigma_{t}(1-\beta^{2})}}
          \end{equation}

It is convenient to describe the conductivity asymmetry by a
parameter $\beta$ such that $\sigma_{\alpha}= \sigma_{t}
(1+\beta)/2$ and $\sigma_{\gamma}=\sigma_{t}(1-\beta)/2$. The
out-of-equilibrium contribution to the resistance then takes the
following form:

\begin{equation}
R^{ne} = \frac{1}{2} \frac{(\beta_{I} -
\beta_{II})^{2}}{\sqrt{eL^{I}\sigma_{t}^{I}(1-\beta_{I}^{2})}+
\sqrt{eL^{II}\sigma_{t}^{II}(1-\beta_{II}^{2})}} \label{Rbeta}
\end{equation}

In the case of the subsystem described by two {\it spin}-channel,
the relaxation $\dot \Psi_{\uparrow \downarrow}$ leads to a {\it
spin-accumulation} effect $\Delta \mu_{\uparrow \downarrow}$ at the
interface of a two identical ferromagnet with antiparallel
configuration. The corresponding resistance contribution is:

\begin{equation}
R_{sa}^{\uparrow \downarrow } =  \frac{ \beta_{s}^{2}}{\sigma_{t}
(1-\beta_{s}^{2})} \, l_{sf} =
        \frac{\beta_{s}^{2}}{\sqrt{eL\sigma_{t}(1-\beta_{s}^{2})}}
        \label{RGMR}
\end{equation}

This expression is the well-known giant magnetoresistance contribution
\cite{Johnson,Wyder,Valet,Levy,PRBThermo,Jedema2,George}.

\subsection{The four channel approximation}
\label{subsec:FourChan}

In the previous subsections, two different electronic relaxation
mechanisms have been invoked separately in order to describe giant
magnetoresistance or anisotropic magnetoresistance.
It is clear however that the two relaxations would take place in
parallel, leading to a more complex redistribution of spins within
the different channels. In the present subsection, we consider a
system in which the two mechanisms coexist, leading to a four
channel model \cite{FourChan}.

\begin{figure}[h!t]
         \begin{center}
         \begin{tabular}{c}
         \includegraphics[height=8cm]{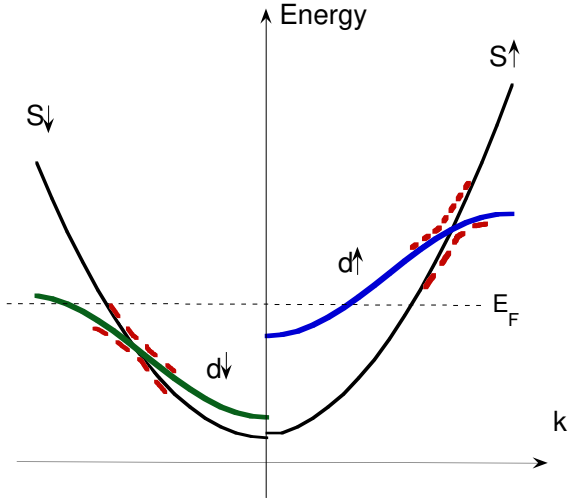}
         \end{tabular}
         \end{center}
         \caption[Band structure]
{ \label{fig:Bands} Generic band structure for a 3d ferromagnet with
s and d bands schematized for an arbitrary direction of the wave
vector k.  The shift between the two d bands for the two spin
carriers up and down is exemplified.  The hybridized zone is
schematized with dotted lines at the junction between s and d bands.
At the Fermi level four different electronic populations can be identified.}
         \end{figure}

     The generic band structure (energy as a function of wave vector
     $\vec{k}$  for a
      given direction) of a 3d ferromagnet is schematized in Fig.
\ref{fig:Bands}.
      The band s is parabolic and the exchange splitting is very small.  In
      contrast, the d bands are strongly shifted between up and down spin
      carriers.  The hybridized zone is schematized by the dotted lines at
      the intersection.

The system is composed by the reservoirs of the injected $s$
electrons and the ferromagnetic layer composed by the $d$ electrons.
At the interface, current injection leads to a redistribution of the
different electronic populations that are governed by spin
polarization and charge conservation laws.  Let us assume that the
current injected is spin polarized in the down polarization
($\downarrow$).  The conservation laws should be written by taking
into account the reaction mechanisms between the different
populations. At short time scales (electronic scattering) the
relaxation channels are assumed to be the
following four \\ \\ \\
        (I) $e_{s \downarrow} \to e_{d \downarrow}$ (spin-conserved $s$-$d$
scattering) \\
        (II) $e_{s \downarrow} \to e_{s \uparrow}$ (spin-flip scattering for
        the $s$ population) \\
        (III) $e_{s \downarrow} \to e_{d \uparrow}$ (spin-flip
        $s$-$d$ scattering) \\
        (IV) $e_{d \downarrow} \to e_{d \uparrow}$ (spin-flip
        scattering for the $d$ population)

        Process (I) is assumed to be the main mechanism responsible for
        anisotropic magnetoresistance (AMR).  Process (II) leads to the
        well-known spin-accumulation effect and was also described in detail in
        the first subsections.  {\it According to the fact that the
        majority-spin $d$ band is full} and lies at a sizable energy below
        the Fermi level, the current $J_{d \uparrow}$ is negligible and the
        channel $d \uparrow $ is frozen.  Processes (III) and (IV) are
        hence negligible \cite{Drouhin}.  Consequently, we are dealing with
        a three-channel model $\{ s\uparrow, s \downarrow, d \downarrow \}
        $.

        The total current $J_{t}$ is composed by the three currents for
each
        channel :
        $J_{t} = J_{s \uparrow}+J_{s \downarrow}+J_{d
        \downarrow}$.
        In order to write the conservation laws, the
        relaxation rate $\dot \Psi_{sd} $, is introduced to
account for $s-d$ spin-conserved scattering, and the relaxation rate
$\dot \Psi_{s} $, is introduced in order to account for spin-flip
scattering.
        {\it Assuming that all channels are in a steady state} (this
        condition will relax in the last section, where the magnetic
        system is coupled to the channels $d \downarrow$) :

\begin{equation}
\left\{
\begin{aligned}
         \frac{\partial n_{t}}{\partial t}& =  - \frac{\partial
J_{t}}{\partial z} = 0 \\
\frac{\partial n_{s \uparrow}}{\partial t}\, &= \,-\frac{\partial J_{s
\uparrow}}{\partial
z} - \, \dot{\Psi}_{s} = 0 \\
\frac{\partial n_{s \downarrow}}{\partial t}\,& = \,-\frac{\partial J_{s
\downarrow}}{\partial z} - \, \dot{\Psi}_{sd}
        + \, \dot{\Psi}_{s} = 0\\
        \frac{\partial n_{d \downarrow}}{\partial t}\,&  =  -\frac{\partial J_{d
\downarrow}}{\partial z} + \dot{\Psi}_{sd} = 0
\end{aligned}
\right. \label{con}
\end{equation}
where $n_{t}, n_{s \uparrow}, n_{s \downarrow}, n_{d \downarrow}$
are respectively the total densities of particles and the density of
particles in the  in the $s \uparrow$, $s \downarrow$, $d
\downarrow$ channels. The system is described by the number of
electrons present in each channel at a given time, that defines the
four currents, plus the entropy of the system.  The conjugate
(intensive) variables are the chemical potentials $\{
\mu_{s\uparrow}, \mu_{s \downarrow}, \mu_{d \uparrow}, \mu_{d
\downarrow }\}$. As described in Appendix B, the application of the
first and second laws of thermodynamics allows us to deduce the
Onsager relations of the system :

\begin{equation}
\left\{
\begin{aligned}
J_{s \downarrow} &= -\frac{\sigma_{s \downarrow}}{e} \frac{\partial
\mu_{s \downarrow}}{\partial z}\\
J_{s \uparrow} &= -\frac{\sigma_{s \uparrow}}{e} \frac{\partial
\mu_{s
\uparrow}}{\partial z}\\
J_{d \downarrow}& = -\frac{\sigma_{d \downarrow}}{e} \frac{\partial
\mu_{d \downarrow}}{\partial z}\\
\dot{\Psi}_{sd} &= L_{sd} \left ( \mu_{s
\downarrow}-\mu_{d \downarrow} \right ) \\
\dot{\Psi}_{s}& = L_{s} \left ( \mu_{s \uparrow}-\mu_{s  \downarrow}
\right )
\end{aligned}
\right. \label{OnsagerF0}
\end{equation}

where the conductivity of each channel $ \{ \sigma_{s \uparrow},
\sigma_{s \downarrow}, \sigma_{d \uparrow}, \sigma_{d \downarrow}
\}$ has been introduced. The first four equations are nothing but
Ohm's law applied to each channel, and the two last equations
introduce new Onsager transport coefficients (see Appendix B),
$L_{sd \downarrow}$ and $L_{s}$, that respectively describe the
$s-d$ relaxation (I) for minority spins under the action of the
chemical potential difference $\Delta \mu_{\downarrow} = \mu_{s
\downarrow}/2-\mu_{d \downarrow}$ and the spin-flip relaxation (II)
under spin pumping $\Delta \mu_{s} = \mu_{s \uparrow}-\mu_{s
\downarrow}/2$. According to Appendix A, the Onsager coefficients are
proportional to the corresponding relaxation times.

For convenience, we define the usual charge
current $J_{0s}= J_{s \uparrow} + J_{s \downarrow}$, the
minority-spin current  $ J_{0 \downarrow} = J_{s \downarrow} + J_{d
\downarrow}$, and the two {\it polarized currents} $ \delta
J_{\downarrow} = J_{s \downarrow} - J_{d \downarrow}$ and $\delta
J_{s}= J_{s \uparrow} - J_{s \downarrow}$. We introduce the
$\sigma_{s}$ and $\sigma_{\uparrow}$ conductivities $ \{ \sigma_{s}
= \sigma_{s \uparrow} + \sigma_{s \downarrow} $ and
$\sigma_{\downarrow} = \sigma_{s \downarrow} + \sigma_{d \downarrow}
\}$. The conductivity imbalance $ \beta_{ \downarrow} $ and $
\beta_{s} $ between respectively the $s \downarrow$ and $d
\downarrow$ channels and the $s \uparrow$ and $s \downarrow$
channels are:

\begin{equation}
\left\{
\begin{aligned}
\beta_{\downarrow}& = \frac{ \sigma_{s \downarrow}-\sigma_{d
\downarrow}}{\sigma_{\downarrow} }\\
\beta_{s}& = \frac{ \sigma_{s \uparrow}-\sigma_{s
\downarrow}}{\sigma_{s}}
\end{aligned}
\right. \label{beta}
\end{equation}
Eqs.~(\ref{con}) becomes :
\begin{equation}
\left\{
\begin{aligned}
\frac{\partial J_{t}}{\partial z}& =  \frac{\partial J_{ d
\downarrow}}{\partial z} + \frac{\partial J_{s}}{\partial z} = \, 0
\\
\frac{\partial J_{0 \downarrow}}{\partial z}& =  \, \dot{\psi}_{s} \\
\frac{\partial \delta J_{\downarrow}}{\partial z}\,& = \, -2
\dot{\psi}_{sd} - \dot{\psi}_{s} \\
\frac{\partial J_{0s}}{\partial z}& =  \, - \dot{\psi}_{sd} \\
\frac{\partial \delta J_{s}}{\partial z}\,& = \, \dot{\psi}_{sd} - 2
\dot{\psi}_{s} \label{cont}
\end{aligned}
\right.
\end{equation}
and, defining the quasi-chemical potentials $\mu_{s} = \mu_{s
\uparrow}+\mu_{s \downarrow}/2$ and  $\mu_{\downarrow} = \mu_{s
\downarrow}/2+\mu_{d \downarrow}$, Eqs.~(\ref{OnsagerF0}) becomes :

\begin{equation}
\left\{
\begin{aligned}
J_{0 \downarrow} &= -\frac{\sigma_{\downarrow}}{2e} \left (
\frac{\partial \mu_{\downarrow} }{\partial z} + \beta_{\downarrow}
\frac{\partial \Delta \mu_{\downarrow}}{\partial
z}  \right ) \\
\delta J_{\downarrow} &= -\frac{\sigma_{\downarrow}}{2e} \left (
\beta_{\downarrow} \frac{\partial \mu_{\downarrow}}{\partial z} +
\frac{\partial \Delta \mu_{\downarrow}}{\partial
z} \right ) \\
J_{0s}& = -\frac{\sigma_{s}}{2e} \left ( \frac{\partial \mu_{s}
}{\partial z} + \beta_{s} \frac{\partial \Delta \mu_{s}}{\partial
z}  \right ) \\
\delta J_{s}& = -\frac{\sigma_{s}}{2e} \left (   \beta_{s}
\frac{\partial \mu_{s}}{\partial z} + \frac{\partial \Delta
\mu_{s}}{\partial
z} \right ) \\
\dot{\Psi}_{sd} &= L_{sd} \Delta \mu_{\downarrow} \\
\dot{\Psi}_{s} &= L_{s} \Delta \mu_{s}
\end{aligned}
\right.
\label{Onsager1}
\end{equation}
The equations of conservation [Eqs.~(\ref{cont})] and the above
Onsager equations lead to the two coupled diffusion equations :

\begin{equation}
\left\{
\begin{aligned}
\frac{\partial^2 \Delta \mu_{\downarrow}}{\partial z^2}\,&=\,
\frac{1}{l_{sd}^2} \, \Delta
\mu_{\downarrow}- \frac{1}{\lambda_{s}^{2}} \Delta \mu_{s} \\
\frac{\partial^2 \Delta \mu_{s}}{\partial z^2}\,&=\,
\frac{1}{\lambda_{sd}^2} \, \Delta \mu_{\downarrow}-
\frac{1}{l_{sf}^2} \, \Delta \mu_{s}
\end{aligned}
      \right.
      \label{diff}
\end{equation}
where

\begin{equation}
\left\{
\begin{aligned}
l_{sd}&\,\equiv \, \sqrt{\frac{ \sigma_{\downarrow}
\left ( 1 - \beta_{\downarrow}^{2}\right )}{4 \, eL_{sd}}} \\
\lambda_{s}& \equiv \, \sqrt{\frac{ \sigma_{\downarrow}
\left ( 1 + \beta_{\downarrow} \right )}{2 \, eL_{s}}} \\
l_{sf}&\,\equiv \, \sqrt{\frac{ \sigma_{s} \left ( 1 -
\beta_{s}^{2}\right
)}{4 \, eL_{s}}} \\
\lambda_{sd}&\,\equiv \, \sqrt{\frac{ \sigma_{s} \left ( 1 -
\beta_{s} \right )}{2 \, eL_{sd}}}
\end{aligned}
\right.
\end{equation}

A solution of Eqs. (\ref{diff}) is

\begin{equation}
\left\{
\begin{aligned}
\Delta \mu_{\downarrow}\,&=\, \Delta \mu_{1} + \Delta \mu_{2} \\
\Delta \mu_{s}\,&=\, \lambda_{s}^{2} \left ( \left (
\frac{1}{l_{sd}^{2}} - \frac{1}{\Lambda_{+}^{2}} \right) \, \Delta
\mu_{1} +  \left ( \frac{1}{l_{sd}^{2}} - \frac{1}{\Lambda_{-}^{2}}
\right ) \Delta \mu_{2}  \right )
\end{aligned}
\right. \label{soldiff}
\end{equation}
with
\begin{equation}
\left\{
\begin{aligned}
\Delta \mu_{1} \,=\, a_{1} e^{\frac{z}{\Lambda_{+}}} + a_{2} e^{-
\frac{z}{\Lambda_{+}}} \\
\Delta \mu_{2} \,=\, b_{1} e^{ \frac{z}{\Lambda_{-}}} + b_{2} e^{-
\frac{z}{\Lambda_{-}}}
\end{aligned}
\right. \label{delta}
\end{equation}
where
$$\Lambda^{-2}_{\pm} = \frac{1}{2}(l_{sd}^{-2}+l_{sf}^{-2})
\left ( 1 \pm \sqrt{1-4 \frac{ l_{sd}^{-2}l_{sf}^{-2}
-\lambda^{-2}_{s}\lambda_{sd}^{-2}}{ \left ( l_{sd}^{-2}+l_{sf}^{-2}
\right )^{2} }} \right ) $$

The constants $a_{1}$, $a_{2}$, $b_{1}$, $b_{2}$ are defined by the
boundary conditions.
        It can then be seen that the usual spin accumulation corresponding
to $\Delta
        \mu_{s}$ also depends on the
        spin-conserved $s-d$ electronic diffusion which is known to be
        efficient \cite{Drouhin} and, conversely, that
        spin-conserved diffusion is able to lead to a spin accumulation,
        or {\it $d$ spin-accumulation} effects.  Accordingly, we
        expect to measure some typical effects related to spin-accumulation
        in single magnetic layers, or if $\beta_{s} = 0$ : this point will
be
        illustrated in the new expression of the magnetoresistance
        (Eq. (\ref{GMR}) below), and in Section IV through the effect of
        current induced magnetization switching (CIMS).  $s-d$
relaxation adds a new
        contribution to the resistance, which plays the role of an interface
        resistance arising from the diffusive treatment of the band
        mismatch \cite{Gijs,Buttler,Levy0}.

        The resistance produced by the usual spin-accumulation
        contribution, plus the contribution of $s-d$ relaxation, are
        defined (see Eq.  ~ (\ref{Res})) by

        \begin{equation}
           R_{sa} = \frac{-1}{eJ_{t}} \int_{B}^{A}
           \frac{\partial}{\partial z} \left (
           \mu_{i} - \Phi(z) \right ) dz
           \end{equation}

where $\Phi(z)$ is the total electric field and $\mu_{i}$ is one of
        the chemical potentials.  Providing that the total current is $J_{t}
= J_{s
        \uparrow}+J_{s \downarrow}+J_{d \downarrow}$, or

         \begin{equation}
        J_{t} \, = \, -\frac{\sigma_{t}}{e} \frac{\partial}{\partial z} \left (
\frac{\sigma_{d
        \downarrow}}{\sigma_{t}} \,  \mu_{d \downarrow}+  \frac{\sigma_{s
        \downarrow}}{\sigma_{t}} \, \mu_{s \downarrow} + \frac{\sigma_{s
        \uparrow}}{\sigma_{t}} \,  \mu_{s \uparrow} \right )
        \label{current}
        \end{equation}
        The total electric field can also be written (from Eqs.
        (\ref{OnsagerF0}))
        as
        \begin{equation}
        \Phi(z) \, = \frac{J_{t}}{\sigma_{t}}= -\frac{1}{e} \left
        (\frac{\sigma_{d \downarrow}}{\sigma_{t}} \frac{\partial \mu_{d
        \downarrow}}{\partial z} + \frac{\sigma_{s \downarrow}}{\sigma_{t}}
\frac{\Delta
        \mu_{\downarrow}}{\partial z} + \frac{\sigma_{s
        \uparrow}}{\sigma_{t}} \frac{\Delta \mu_{s}}{\partial z} \right )
        \end{equation}
where $\sigma_{t}=\sigma_{s \uparrow} + \sigma_{s \downarrow} +
        \sigma_{d \downarrow}$.  The resistance is given by :

        \begin{equation}
           R_{sa} = - \frac{1}{e J_{t}} \int_{A}^{B} \left (
           \frac{\sigma_{s\downarrow}}{\sigma_{t}} \frac{\partial \Delta
           \mu_{\downarrow}}{\partial z} + \frac{\sigma_{s
           \uparrow}}{\sigma_{t}}\frac{\partial \Delta
           \mu_{s}}{\partial z} \right )dz
           \label{GMR}
\end{equation}

This three-channel model brings to light the interplay between band
mismatch effects and spin accumulation, in a diffusive approach.  It
is interesting to note that the local neutrality charge condition
which is often used (see for instance Eq.  (4) in \cite{Rashba}) was
not included, as described in Appendix A. On the contrary, we have
imposed the conservation of the current at any point of the conductor.
Indeed, electron transfer from a channel to another where the electron
mobility is different, induces a local variation of the total current.

The resolution of the coupled diffusion equations is discussed
elsewhere \cite{FourChan}.

\subsection{Domain wall scattering}

    In the description performed until now, the spin quantification axis
        that defines up and down spin states was fixed through the whole
        structure (i.e. through the layers and the interfaces).  Providing
        that the spin quantification axis follows the direction of the
        magnetization, it could be non-uniform throughout a ferromagnetic
layer,
        or crossing an interface.  This is especially the case in the
presence
        of a magnetic domain wall.  In a thin enough magnetic domain wall
the
        spin would not follow adiabatically the quantification axis, leading
        to spin-dependent domain wall scattering (DWS)
\cite{Viret,LevyDW,Ulrich,DWS}.
        This effect has been investigated intensively in the last decades in
        various structures \cite{Marrows}. The
        underlaying idea is however rather simple, and can be
        formulated easily with a generalization of the two-spin channel
        approach.  For the sake of
        simplicity, this generalization will be performed only for the two
        electronic populations $\{\alpha, \gamma \}$.

         As performed in reference \cite{PRBThermo} (and appendix B), we start
         with the conservation of the particles for the two
         channels, in a discreet model.  The system is described by a layer
         $\Sigma_{k}$ in contact with a left layer $\Sigma_{k-1}$ and a
right
         layer $\Sigma_{k+1}$.  The spin-flip scattering introduced in the
         previous sections is described be the reaction rate $\dot
\Psi^{k}$.
         A probability $(1-\Delta \epsilon(k))$ of spin-flip alignment along
         the quantification axis is introduced.  In the case of ballistic
         alignment $(1-\Delta \epsilon(k)) = cos^{2}(\Delta\theta(k)/2)$
         where $\Delta \theta (k)$ is the angle between the magnetization of
         two adjacent layers $\Sigma_{k-1}$ and $\Sigma_{k}$.  The
         conservation of the particles is now describes by:

         \begin{equation}
         \left\{
         \begin{aligned}
         \frac{dN_{\alpha}}{dt} = (1-\Delta \epsilon(k)) I_{\alpha}^{k-1
\rightarrow
         k} -I_{\alpha}^{k \rightarrow  k+1} + \Delta \epsilon(k)
         I_{\gamma}^{k-1 \rightarrow  k} - \dot \Psi^{k} \\
         \frac{dN_{\gamma}}{dt} = (1- \Delta\epsilon(k))
         I_{\gamma}^{k-1 \rightarrow  k}  -I_{\gamma}^{k \rightarrow  k+1} +
         \Delta \epsilon(k) I_{\alpha}^{k-1 \rightarrow
         k} + \dot \Psi^{k}
         \label{contDW}
         \end{aligned}
         \right.
         \end{equation}

         With the notation introduced in the previous sections, the entropy
         variation can be written in the following way (Appendix B):

         \begin{eqnarray}
T\frac{dS}{dt}\,& = &\, P_{\Phi}^{R^l \to 1} - P_{\Phi}^{\Omega \to
R^r} \nonumber\\
& & +\, \sum_{k=2}^{\Omega}\frac{1}{2} \left(\Delta \mu^{k-1}-\Delta
\mu^k + 2(1-\Delta \epsilon(k)) \Delta \mu^{k} \right ) \, \delta
I^{k-1
\to k}_{s}  \nonumber\\
& &  + \sum_{k=2}^{\Omega} \frac{1}{2} (\mu^{k-1} - \, \mu^{k})\,
I^{k-1 \to k}_{0} +
        \sum_{k=1}^{\Omega} \Delta \mu^{k} \, \dot \Psi^{k}
\label{entropytot}
\end{eqnarray}

where we have introduced $I_{0} = I_{\alpha} + I_{\gamma}$, $\delta
I = I_{\alpha} - I_{\gamma}$, $\mu_{0} = \mu_{\alpha} +
\mu_{\gamma}$, and $\Delta \mu = \mu_{\alpha} - \mu_{\gamma}$. The
terms $P_{\Phi}^{R^l \to 1}$ and $P_{\Phi}^{\Omega \to 1}$ stand for
heat and chemical transfer from the reservoirs to the system
$\Sigma$.

After performing the continuum limit, the internal entropy
production $\mathcal I $ (or irreversibility) reads:
\begin{equation}
          T. \mathcal I = - \frac{1}{2}\frac{\partial \mu_{0}}{\partial z}
          J_{0} +
          \frac{1}{2} \left ( - \frac{\partial \Delta \mu }{\partial z} +
          2 \epsilon \Delta \mu \right) \delta J
            + \Delta\mu \dot \Psi
\label{EntropDW}
\end{equation}

The first term is the Joule effect, the second is the dissipation
terms related to the spin-accumulation process that occurs at the
interface, or for magnetic domain wall, and the third term is the
dissipation due to spin-flip (or s-d) electronic relaxation.  The
expression of the entropy production Eq.~(\ref{EntropDW}) allows the
Onsager relations generalizing Eq.~(\ref{Onsager0}) or
Eq.~(\ref{OnsagerF0}) to be deduced:

\begin{equation}
\left\{
\begin{aligned}
J_{0} &= -\frac{\sigma_{0}}{2e} \left ( \frac{\partial
\mu_{0}}{\partial z} + \beta  \left ( \frac{\partial
\Delta \mu}{\partial z} - 2 \epsilon \Delta \mu  \right ) \right ) \\
\delta J &= -\frac{\sigma_{0}}{2e} \left ( \beta  \frac{\partial
\mu_{0}}{\partial z} + \frac{\partial
\Delta \mu}{\partial z} - 2 \epsilon  \Delta \mu \right ) \\
\dot \Psi& = L_{\alpha \gamma} \Delta \mu
\end{aligned}
\right. \label{OnsagerDW}
\end{equation}
where $\epsilon = lim_{\Delta k \rightarrow 0} \frac{\Delta
\epsilon(k)}{\Delta k}$, and, as already introduced: $\sigma_{0} =
\sigma_{\alpha} + \sigma_{\gamma}$ and $\beta = (\sigma_{\alpha} -
\sigma_{\beta})/\sigma_{t}$.

The diffusion equation for $\Delta \mu$, obtained in the stationary
regime, is modified accordingly:

\begin{equation}
\frac{\partial^{2} \Delta \mu}{\partial z^{2} } = \left (
\frac{1}{l_{diff}^{2}} + \frac{1}{l_{DW}^{2}} \right ) \Delta \mu +
\frac{1}{\kappa} \frac{\partial \Delta \mu}{\partial z}
          \end{equation}
where the length $l_{diff}$ as been defined in the first section
Eq.~(\ref{ldiff}) :

\begin{equation}
          l_{diff}= \sqrt{\frac{\sigma_{0}(1-\beta^{2})}{2eL}}
\end{equation}
the domain wall diffusion length $l_{DW}$ is defined as:
\begin{equation}
l_{DW} = \sqrt{\frac{(1-\beta^{2})}{4 \epsilon}}
\end{equation}
while the length $\kappa$ is given by:
\begin{equation}
\kappa^{-1} = \epsilon \frac{2 \beta^{2}}{1-\beta^{2}}
\end{equation}
The magnetoresistance is modified with respect to
Eq.~(\ref{DiffEq}), due to the new term $\partial_{z}\Delta \mu(z)$
in the diffusion equation.  It is worth pointing out that a spin
accumulation $\Delta \mu (z) \ne cst$ should be expected in case of
spin polarized current ($\beta \ne 0$) even without the usual
spin-flip contribution, i.e. in the ballistic limit.

\section{Ferromagnetic brownian motion and magnetization switching}

\subsection{Thermokinetic derivation of the Fokker-Planck equation}

     The description performed in the previous
sections is related to the transport properties of charge carriers
in case of spin polarized current.  In spintronics experiments, the
electric current is spin-polarized through a ferromagnetic layer,
but it is not necessary to describe the ferromagnetic order
parameter as such. This is of course no longer the case for current
induced magnetization switching experiments, where the magnetization
is the measured variable.

The magnetization is a fascinating degree of freedom, that has to be
described in length in terms of rotational Brownian motion.  The
description of the dynamics of ferromagnetic particles coupled to a
heat bath is a very active field of investigation
\cite{Neel,Brown,Brown2,Coffey,Palacios}, and the resulting
predictions are rather well known and validated experimentally at
large \cite{WW,WW2,Fruch} and short
(\cite{Bertram,Smith,Chappert,ChappertBal,Russek}) time scales.  The
magnetization relaxation described here is limited to the so-called
N\'eel relaxation that involves only the magnetic moment, in
contrast to the Debye inertial relaxation occurring in ferrofluids
(in which the ferromagnetic particles rotates in a viscous
environment, leading to surprising inertial effects like negative
viscosity \cite{Rubi2}).

The aim of this subsection is first to show that the rotational
Fokker-Planck equation governing the dynamics of the magnetization
$\vec{M}$ of one monodomain particle coupled to the heat bath can
also be obtained applying step by step the approach used in the
previous sections.  The resulting Fokker-Planck equation with the
corresponding Onsager transport coefficients, and the
hypothesis performed, can then be compared term by term to the previous
study of spin-dependent charge transport.

\begin{figure}[h!]
\begin{center}
     {\includegraphics[height=5cm]{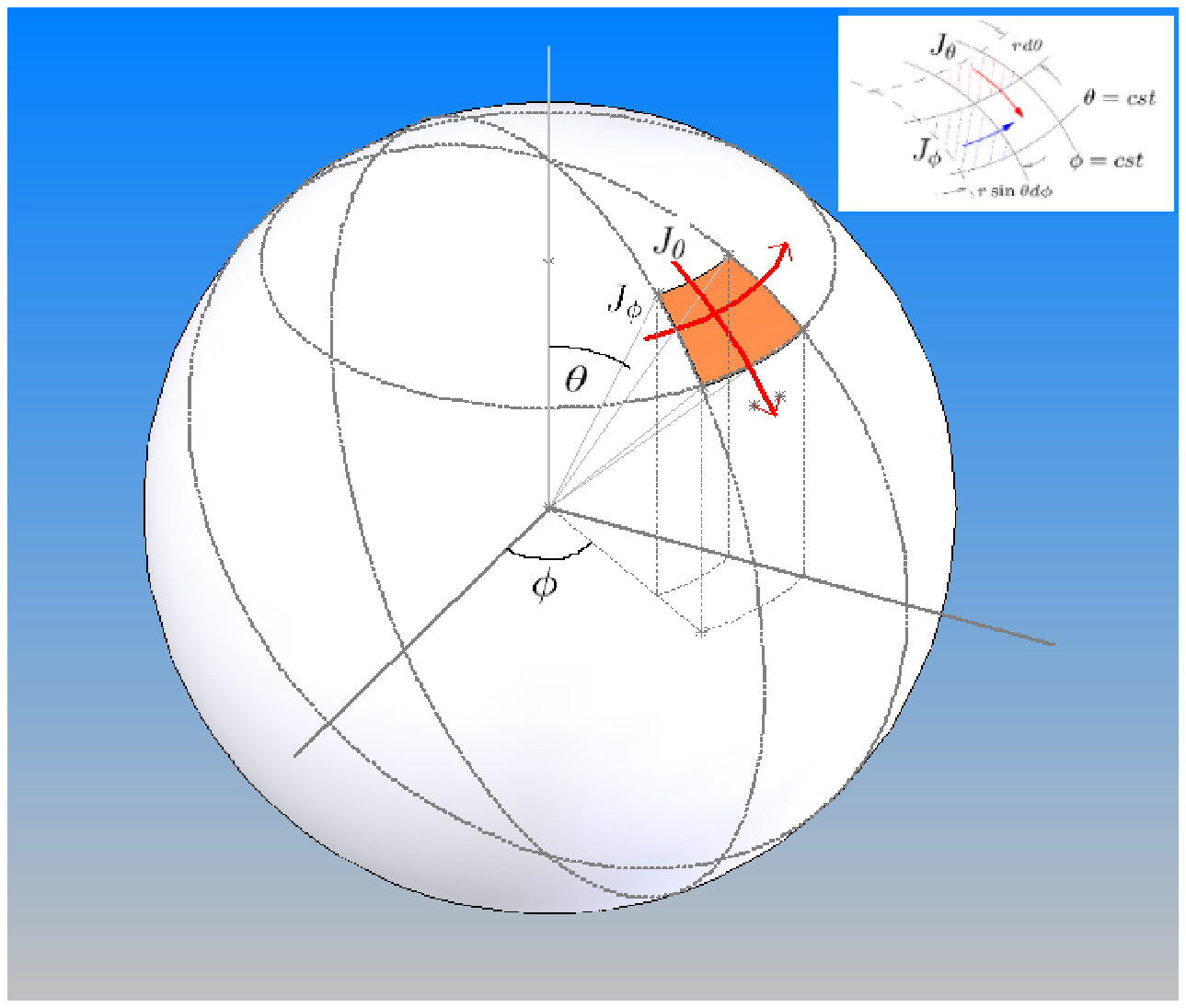}}
     \qquad \qquad
     {\includegraphics[height=5cm]{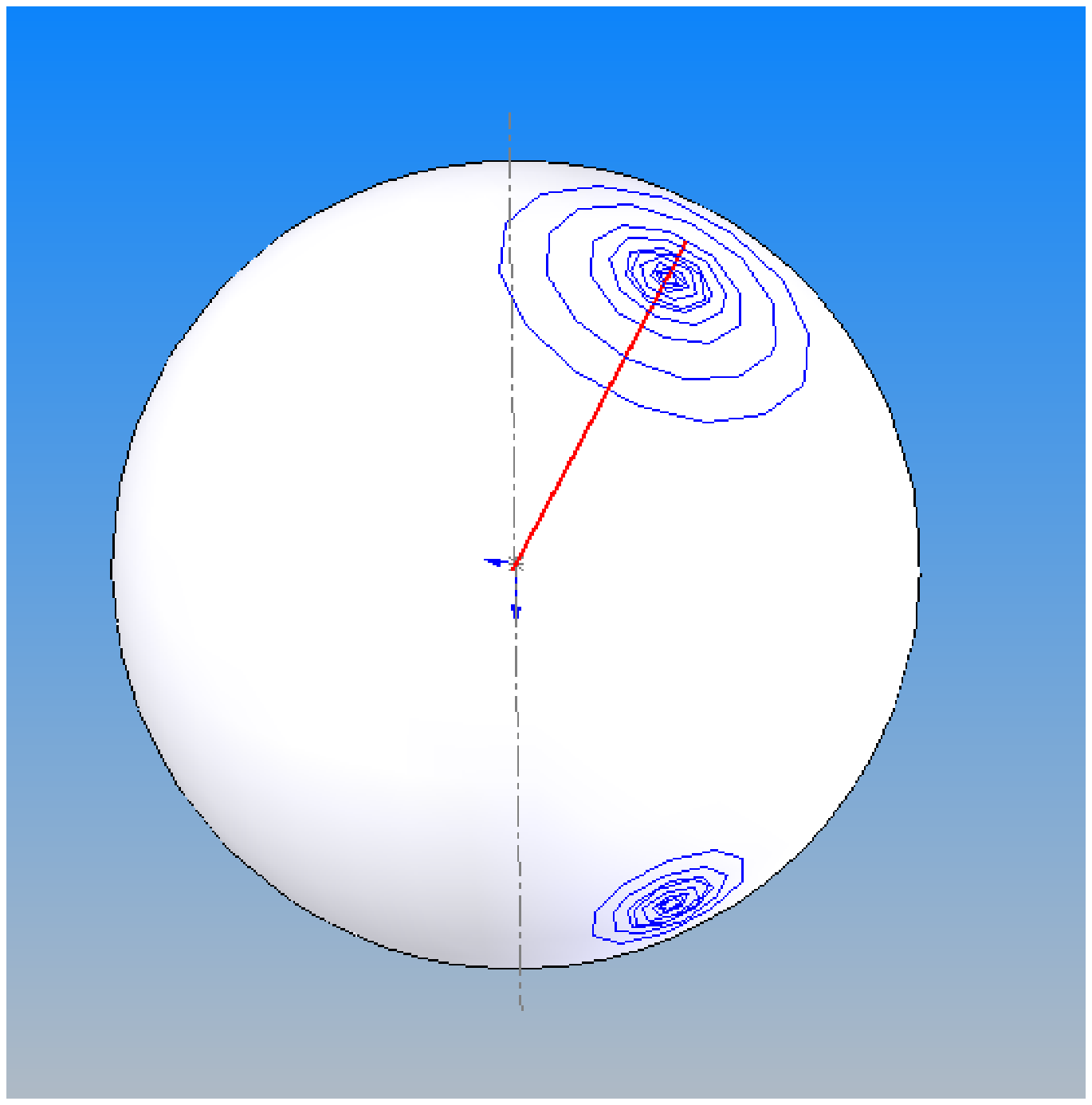}}
\end{center}
\caption[The sphere of representative points] {a) The figure from
the left illustrates the flow of representative points over the unit
sphere: $J^{\theta}$ and $J^{\phi}$. b) The figure from the right
illustrates a particular case of distribution of points on the
sphere: the points are concentrated at two attractors, one with more
particles than the other (asymmetric double well potential).}
\label{sphere}
\end{figure}

\subsubsection{ Geometrical representation of the statistical
ensemble}

\ \ \ \ Let $\Sigma$ be a statistical ensemble of $N$ identical
monodomain particles of volume $v$, having the same energy per unit
volume $V^{mag}(\theta,\phi)$, magnetization $\vec{M}$
and thermostat temperature $T$. The vector $\vec{M}$ is defined by
the angles
$\theta$ and $\phi$. The ensemble $\Sigma$ can be
represented by a distribution of representative points over the unit
sphere (fig. \ref{sphere}) with a density $n(\theta,\phi)$.

We divide the ensemble of representative points $\Sigma$ in
sub-ensembles $\Sigma_{\theta,\phi}$ such that
the magnetization is confined within the solid angle $\delta
V_{\theta,\phi}=sin\theta d\theta d\phi$. (i.e. the representative
points lie between two consecutive parallels and meridians over the
sphere).

As the particles undergo changes of magnetization orientation, the
representative points move on the sphere, and there is a net surface
flux of representative points $\vec{J}^{mag}$; the representative
points move from one sub-ensemble $\Sigma_{\theta,\phi}$ to another
sub-ensemble $\Sigma_{\theta+\Delta \theta,\phi+\Delta\phi}$. The
probability of finding a particle with the magnetization orientation
within the solid angle $d V_{\theta,\phi}$ at a given time $t$ is
$dP(\theta,\phi,t)=\frac{n(\theta,\phi,t )}{N}\,d V_{\theta,\phi}$.


\subsubsection{Conservation laws}

The sub-ensembles of representative points $\Sigma_{\theta,\phi}$
are described by the following extensive parameters: the
entropy $dS=s(\theta,\phi,t )d V_{\theta,\phi}$, the
number of points $dN=n(\theta,\phi,t )dV_{\theta,\phi}$
and the energy $dE=e(\theta,\phi,t)d V_{\theta,\phi}$,
where $s$ and $e$ are the entropy and energy densities. The flow (of
points, energy, and entropy) is described by the flux $\vec{J}$
($\vec{J}_{n}, \ \vec{J}_{e}$ and $\vec{J}_{s}$):

\begin{equation}
         \vec{J}=
           J^{\theta}\vec{u}_{\theta}+
           J^{\phi}\vec{u}_{\phi}
\end{equation}
and accounts for the flow of the corresponding magnetic moments
relaxing or precessing along the coordinates $\theta , \phi$, where
$\vec{u}_{r}$, $\vec{u}_{\theta}$, $\vec{u}_{\phi}$ are the
unit vectors in the spherical coordinate system.

The conservation laws of the number, energy and entropy of the
particles contained in the  sub-ensemble $\Sigma_{\theta,\phi}$
write :
\begin{equation}
        \left \{
         \begin{aligned}
           \frac{\partial n}{\partial t}
             &=-\,div \, \vec {J_{n}}\\
           \frac{\partial e}{\partial t}
             &=-\,div \, \vec{J_{e}}\\
           \frac{\partial s}{\partial t}
             &=-\,div \, \vec {J_{s}}+\mathcal{I}
         \end{aligned}
\right. \ \Rightarrow \ \left \{
         \begin{aligned}
           \frac{\partial n}{\partial t}
             &=-\frac{1} {sin \theta}
                \frac{\partial} {\partial \theta}             \left (
                     J_{n}^{\theta}sin\theta
                   \right)
           - \frac{1}{sin \theta}
             \frac{\partial J_{n}^{\phi}}
                  {\partial \phi}     \\
           \frac{\partial e}{\partial t}
             &=-\frac{1} {sin \theta}
                \frac{\partial} {\partial \theta}             \left (
                     \mathcal{J}_{e}^{\theta}sin\theta
                   \right)
           - \frac{1}{sin \theta}
             \frac{\partial \mathcal{J}_{e}^{\phi}}
                  {\partial \phi}       \\
              \frac{\partial s}{\partial t}
             &=-\frac{1} {sin \theta}
                \frac{\partial} {\partial \theta}             \left (
                     \mathcal{J}_{s}^{\theta}sin\theta
                   \right)
           - \frac{1}{sin \theta}
             \frac{\partial \mathcal{J}_{s}^{\phi}}
                  {\partial \phi}+\mathcal{I}
             \end{aligned}
\right. \label{Cons}
\end{equation}
\linebreak where in contrast to the energy and number of particles,
the entropy $s$ is not a conservative quantity, and an internal
entropy production term $\mathcal I $ (or $\it irreversibility$) is
added to the entropy flux $\vec{J_{s}}$ (third equation in Eq.
(\ref{Cons})).

The expression of the first law of thermodynamics, allows the energy
variation to be expressed as a function of the partial derivatives
that define the chemical potentials and the temperature $\tilde \mu
\equiv \frac{\partial e}{ \partial n}$ and $T \equiv \frac{\partial
e}{\partial s}$ .  The intensive variables $\tilde \mu$ and $T$ are
also functions of ($\theta,\phi,t$) except if imposed by a reservoir.
\begin{equation}
\frac{\partial s}{\partial t}= \frac{1}{T} \frac{\partial
e}{\partial t} - \frac{\tilde \mu}{T}  \frac{\partial n}{\partial t}
\label{Gibbs}
\end{equation}
where $\tilde \mu \equiv \partial_{n}e$ contains all contributions
to the energy (see below)).

The expression of the internal entropy variation can be obtained
using the conservation laws:

\begin{align}
\frac{\partial s}{\partial t}=
        -\frac{1}{T}\,div \, \vec{J_{e}}
        +\frac{\tilde \mu}{T}\,div \, \vec{J_{n}}
\label{condiv}
\end{align}

or, in an other form:

\begin{align}
      \frac{\partial s}{\partial t}
        =-div
            \left(
              \frac{1}{T}\,\vec{J_{e}}
             -\frac{\tilde\mu}{T}\,\vec{J_{n}}
            \right)
         +\vec{J_{e}}\cdot\vec{grad}
            \left(
              \frac{1}{T}
            \right)
         -\vec{J_{n}}\cdot\vec{grad}
           \left(
             \frac{\tilde\mu}{T}
           \right)
\label{sep}
\end{align}

Comparing this last equation with the third eq. from (\ref{Cons}),
we can deduce the form of the entropy production $\mathcal{I}$:

\begin{align}
      \left\{
       \begin{aligned}
        \vec{J_{s}}=&\frac{1}{T}\,\vec{J_{e}}
             -\frac{\tilde\mu}{T}\,\vec{J_{n}}\\
        \mathcal{I}=&\,\vec{J_{e}}\cdot\vec{grad}
            \left(
              \frac{1}{T}
            \right)
         -\vec{J_{n}}\cdot \vec{grad}
           \left(
             \frac{\tilde\mu}{T}
           \right)
       \end{aligned}
      \label{SProd}
      \right.
\end{align}

The entropy production $\mathcal{I}$ is a sum of products between
      the fluxes $\vec{J_{k}}$ and the corresponding
      conjugate forces $\vec{F_k}$  \cite{Mazur}.

We assume in the following that the temperature $T(\theta, \phi)=T$
is fixed by a unique thermostat: the first term in the right hand side of
the eq. (\ref{SProd}) vanishes.

      A sufficient condition to impose the second
law of thermodynamics $\mathcal I \ge 0$ is then to build a
quadratic form. This leads us to define the matrix $\mathcal{L}$ of Onsager
transport coefficients $L_{i j} (\theta,\phi)$ (that are state
functions of dimension $[energy]^{-1}[time]^{-1}$) such that $J_{i}=
\Sigma_{j} \left (L_{i j} \partial_{j} \tilde \mu \right )$, where
the symmetrized $\mathcal{L}$ matrix is positive.  The {\it Onsager
reciprocity relations} impose furthermore that $L_{i j} = \pm L_{j
i}$, where the sign (-) is present if $L_{i j}$ is a function of the
magnetic field (there is no angular velocity here) \cite{DeGroot}.

The following relations are deduced:
\begin{gather}
      \vec{J_n}=-\mathcal{L}\ \vec{grad}\,\tilde\mu
      \quad \Rightarrow \quad
      \left\{
       \begin{aligned}
          J_{n}^{\theta}\,
          &=
             -L_{\theta\theta }
              \frac{\partial \tilde \mu}{\partial \theta}
             -L_{\theta \phi }
              \frac{1}{sin\theta}
              \frac{\partial \tilde \mu}{\partial \phi}
            \\
          J_{n}^{\phi} \,
          &=
             -L_{\phi \theta} \frac{\partial \tilde \mu}{\partial \theta}
             -L_{\phi \phi}
              \frac{1}{sin\theta}
              \frac{\partial \tilde \mu}{\partial \phi}
       \end{aligned}
      \right. \label{flux}
\end{gather}

where $L_{\theta \phi}= - L_{\phi \theta}$. The first of equations
(\ref{Cons}) re-writes:
      \begin{equation}
       \frac{\partial n_{(\theta,\phi,t)}} {\partial t}
        =-\vec{div\ J_n}
        =+\vec{div}
            \left(
              \vec{L}\
              \vec{grad} \tilde\mu
            \right) \Rightarrow
       \label{ContMag}
      \end{equation}

\begin{multline}
         \frac{\partial n_{(\theta,\phi,t)}} {\partial t}
           =
           \frac {1} {sin\theta}\,
           \frac  {\partial} {\partial \theta}
           \left[
             sin\theta
             \left(
               L_{\theta\theta }
               \frac{\partial\tilde{\mu}_{(\theta,\phi,t)}}
                    {\partial \theta}
               +
               \frac {1} {sin\theta}
               L_{\theta \phi} \frac{\partial\tilde{\mu}_{(\theta,\phi,t)}}
                    {\partial \phi}
             \right)
          \right] \\
          +
          \frac {1} {sin\theta}
          \frac  {\partial} {\partial \phi}
          \left[
            L_{\phi\theta }
            \frac{\partial \tilde{\mu}_{(\theta,\phi,t)}}
                 {\partial \theta}
            +
            \frac {1} {sin\theta}
            L_{\phi\phi}
            \frac{\partial \tilde{\mu}_{(\theta,\phi,t)}}
                 {\partial \phi}
          \right]
\label{EqGene}
\end{multline}

where $L_{\theta \theta } \ge 0$ and $L_{\phi \phi} \ge 0$.  This is
the general expression of the density variation $\partial_{t}
n(\theta,\phi,t)$ of particles number from the sub-ensemble
$\Sigma_{\theta,\phi}$. Note that there is no
relaxation terms ($\dot \Psi$) in the conservation law
(\ref{ContMag}) of the magnetization: the flow of representative points is
conserved on the unit sphere. This assumption will be removed in the case
of spin injection performed with electric currents (see next section below).

In the same manner as for Eq. (\ref{RelaxTau}), the Onsager
coefficients are related to the relevant relaxation times. In the
case of ferromagnetic insulators, the relaxation channels are well
defined \cite{Sparks}, and the coefficients $L_{ij}$ are directly
related to the relaxation times $T_{1}$ and $T_{2}$ measured in
ferromagnetic resonance experiments.

\subsection{The rotational Fokker-Planck equation}

In thermokinetics, the intensive parameter which controls
the number of particles of a sub-ensemble is the chemical potential
$\tilde \mu$.  The relevant energy terms are contained in the
deterministic potential $\mu$, and the stochastic term is defined by
thermal fluctuations due to the coupling to a relevant heat bath.
Anticipating the last section, it is worth pointing out that the
relevant heat bath is defined by the degrees of freedom of the
environment which are that of the lattice or that of the electronic
system (as discussed below).  The fluctuations are taken into account
through a temperature dependent chemical potential
that takes the following form
(derived in the general case by P. Mazur in Ref. \cite{Mazur}) :

\begin{equation}
         \tilde \mu \equiv k_{B}T \ln \left
         (\frac{n}{N}\right ) + vV^{mag}(\theta,\phi)+\mu_0
\label{mu}
\end{equation}

The first term in the right hand side of Eq. (\ref{mu}) is
responsible for thermal agitation at temperature $T$, the second
term $vV^{mag}$ represents the magnetic energy of one particle that
defines the local magnetic field
$\vec{H_{eff}}=-\partial_{\vec{M}} (V^{mag})$ and the third
term is a constant which is related to the chemical nature of the
particles.

The local equilibrium condition $\partial_{i} \tilde \mu =0$
defines stationary flux (due to both drift and diffusion) that are mutually
compensated along the coordinate $i$. This
point is well illustrated in the work of Guggenheim while
introducing the electro-chemical potential \cite{Guggenheim} in
order to generalize the description of an electric fluid to ionic
solutions.

Inserting the expression of $\tilde{\mu}$, and using the reciprocal
relation $L_{\theta \phi}=-L_{\phi \theta}$, the equations for
fluxes and the variation of particles take the form:

\begin{gather}
\left\{
\begin{aligned}
          J_{n}^{\theta}\,
          &=-\left (
                 h'
                 \frac{\partial V^{mag}}{\partial \theta}
                -\frac{g' }{sin\theta}
                 \frac{\partial V^{mag}}{\partial \phi}
              \right ) n
             -\left (
                h'\frac{k_BT}{v}
                \frac{\partial n}{\partial \theta}
               -
                \frac{g'}{sin\theta}
                \frac{k_BT}{v}
                \frac{\partial n}{\partial \phi}
              \right )
            \\
          J_{n}^{\phi} \,
          &=-\left (
                g'
                \frac{\partial V^{mag}}{\partial \theta}
               +
                \frac{k'}{sin\theta}
                \frac{v}{k_BT}
                \frac{\partial V^{mag}}{\partial \phi}
              \right ) n
             -\left (
                g'
                 \frac{k_BT}{v}
                 \frac{\partial n}{\partial \theta}
                +
                 \frac{k'}{sin\theta}
                 \frac{\partial n}{\partial \phi}
              \right)
\end{aligned}
\right. \label{fluxnous}
\end{gather}
where the following standard notations have been introduced:
\begin{equation}
         h'=\frac{L_{\theta\theta } v}{n} \ge 0; \ \ \ \ \ \ \ \
         g' =-\frac{L_{\theta \phi } v}{n}=
              \frac{L_{\phi \theta } v}{n};\ \ \ \ \ \ \ \
         k' = L_{\phi \phi} \frac{k_BT}{n} \ge 0
\end{equation}

Assuming that $g'$ is constant, Equation Eq. (\ref{EqGene})
rewrites:

\begin{multline}
         \frac{\partial n(\theta,\phi)} {\partial t}=
           \frac {1} {sin\theta}\,
           \frac  {\partial} {\partial \theta}
           \left\{ sin\theta
               \left[
                \left (
                 h'
                 \frac{\partial V^{mag}}{\partial \theta}
                -\frac{g' }{sin\theta}
                 \frac{\partial V^{mag}}{\partial \phi}
              \right ) n
             + h'\frac{k_BT}{v}
               \frac{\partial n}{\partial \theta}
             \right]
          \right \}\\
        + \frac{1}{sin\theta}
          \frac  {\partial} {\partial \phi}
           \left \{
          \left (
                g'
                \frac{\partial V^{mag}}{\partial \theta}
               +
                \frac{k'}{sin\theta}
                \frac{v}{k_BT}
                \frac{\partial V^{mag}}{\partial \phi}
              \right ) n
             + \frac{k'}{sin\theta}
               \frac{\partial n}{\partial \phi}
           \right \}
\label{FPEnous}
\end{multline}

The expression Eq. (\ref{FPEnous}) represents the rotational
Fokker-Planck equation obtained by thermokinetic means. Its
expression is identical to that obtained by Brown \cite{Brown}
through stochastic calculations.

Furthermore, the Onsager matrix also follows the symmetry of the
system, and is invariant by rotation around the anisotropy axis, so
that :$ L_{\phi,\phi} = L_{\theta,\theta}$.  The following
relations are obtained :
\begin{equation}
         L_{\theta\theta}=L_{\phi\phi}=\frac{h'n}{v}=\frac{k'n}{k_BT}
\end{equation}

which permits us to write the equations above in a compact
vectorial form:
\begin{equation}
\left \{
\begin{aligned}
        \vec{J}&=
         -g' \vec{u}_{r}\times
                \left(
                  n \vec \nabla V^{mag}+\frac{k_BT}{v} \vec \nabla n
                \right)
         +h'\vec{u}_{r}\times
           \left[
             \vec{u}_{r}\times
             \left(
               n \vec \nabla V^{mag}+\frac{k_BT}{v} \vec \nabla n
             \right)
           \right] \\
\frac{\partial n}{\partial t}&=
            g'\vec \nabla
            \left \{
               \vec{u}_{r}\times
                \left(
                  n \vec \nabla V^{mag}+\frac{k_BT}{v}\vec \nabla n
                \right)
            \right \}
             - h'\vec \nabla
               \left \{
                \vec{u}_{r}\times
                 \left[
                 \vec{u}_{r}\times
                   \left(
                     n \vec \nabla V^{mag}+\frac{k_BT}{v} \vec \nabla n
                   \right)
                \right]
            \right \}
\end{aligned}
\right .
\label{rotFPE}
\end{equation}

where the gradient $\vec \nabla$ is in spherical coordinates
and $\vec{u}_{r}$ is the spherical radial unit vector.
\smallskip
It is to be noticed that the second equation has drift terms which
contain $\nabla V^{mag}$, and diffusion terms which contain $\vec \nabla
n$. The terms $k'=h'\frac{k_BT}{v}$ and $g'\frac{k_BT}{v}$ are the
rotational diffusion coefficients and the terms $g'n$ and $h'n$
represent drift coefficients.

\subsection{Landau-Lifshitz-Gilbert equation with diffusion}

Furthermore, using the first equation from (\ref{rotFPE}),
      one can deduce the Landau-Lifshitz-Gilbert equation with
      diffusion. As $\vec{J}=n\frac{d \vec{u}_{r}}{d t}$, we arrive at
the equation:

\begin{equation}
         \frac{d \vec{u}_r}{d t}=
             -g' \vec{u}_r\times
                \left(
                  \vec \nabla V^{mag}+\frac{k_BT}{v}
                  \frac{\vec \nabla n}{n}
                \right)
         +h'\vec{u}_r\times
           \left[
             \vec{u}_r\times
             \left(
               \vec \nabla V^{mag}+\frac{k_BT}{v}
               \frac{\vec \nabla n}{n}
             \right)
           \right]
      \label{LLG}
\end{equation}

where the first term in the right hand side is {\it the precession}
term, and the second term in the right hand side describes the {\it
longitudinal
relaxation}.
      Multiplied by the amplitude of the magnetization $M_s$,
becomes the \textbf{Landau-Lifshitz-Gilbert equation} with diffusive
terms:
\begin{equation}
\frac{d \vec{M}}{d t}=
               g'M_s\vec{M}\times
              \left(
                \vec{H_{eff}}
              - \frac{k_BT}{vM_s}
                \frac{\vec \nabla n}{n}
              \right)
         +h'\left[
              \vec{M}\times
              \left(
              \vec{H_{eff}}
             -
              \frac{k_BT}{vM_s}
              \frac{\vec \nabla n}{n}
              \right)
            \right]
           \times\vec{M}
\end{equation}

Experimentally, the first contribution can be observed through
ferromagnetic resonance measurements (FMR) at typical frequencies of
tens of GHz (around 100 psec time scales).  The thermalization time
(proportional to 1/g' ; see next section)  is given by the width of the
resonance peaks.  Both frequency and time resolved noise experiments
have been also performed in order to measure precession and thermal
spin-waves \cite{Bertram,Smith,Chappert,ChappertBal,Russek}.  In the above
measurements, the data are averaged over many shots (or
trajectories) near an equilibrium position (linear response regime).
In contrast, the measurements at large time scales, typically beyond
few nanosecond, access the magnetization reversal for
which {\it the precession terms can be neglected}.  The {\it one
shot measurements} give a direct access to the stochastic nature of
the signal \cite{WW,WW2,Guittienne,MSU,SPIE}: a snapshot is a
statistical event, namely the magnetization reversal
from one metastable state to the other, that is governed by the random
fluctuations, described by a "Langevin force" that is not present in
the averaged LLG equation.

Equation
Eq.  (\ref{LLG}) can be put into the Gilbert form by performing the
cross product $\times \vec M$ at the left and right hand side of
the equation.  We obtain the well-known Gilbert equation, that defines
the Gilbert damping parameter $\eta $:

\begin{equation}
         \frac{d \vec{M}}{dt}=
             \Gamma \vec{M} \times
                \left ( \vec H_{eff} - \eta \frac{d \vec{M}}{dt} \right)
      \label{Gilbert}
\end{equation}

where $\Gamma$ is the gyromagnetic factor.  The constant h', g' and k'
are related to the Gilbert damping coefficient $\eta$, $\Gamma $ and
the magnetization at saturation $M_{s}$ through the following
relations:

\begin{equation}
         g' = \frac{\Gamma}{ \left (1+ (\eta \Gamma M_{s} )^{2} \right
         )M_{s}} \ge 0
        ;\ \ \ \ \ \ \ \
        h' = \eta \Gamma M_{s} g'; \ \ \ \ \ \ \ \
\end{equation}

or
\begin{equation}
      \eta = \frac{1}{\Gamma \, M_{S}} \sqrt{\frac{\Gamma}{g'M_{s}} -1} \ge 0
      \end{equation}

\subsection{Activation regime and N\'{e}el-Brown law}

\subsubsection{Neglecting precession}

In the slow relaxation measurements (the so called magnetic
after-effect), relaxation is governed by activation over a
potential barrier. At longtime scales (beyond tens of nanosecond to
hour),
the precessional terms can be neglected. The expression for the
surface current fluxes Eq. (\ref{fluxnous}) becomes

\begin{equation}
\left\{
\begin{aligned}
      J_{n}^{\theta}\,
          &=-  h'
               \frac{\partial V^{mag}}{\partial \theta}
               \,n
               -h'\frac{k_BT}{v}
                \frac{\partial n}{\partial \theta}
          \\
          J_{n}^{\phi} \,
          &=-   g'
                \frac{\partial V^{mag}}{\partial \theta}
                n
                -  g'\frac{k_BT}{v}
                \frac{\partial n}{\partial \theta}
\label{fluxanisotropy}
\end{aligned}
\right.
\end{equation}
and the corresponding Fokker-Planck equation becomes:
\begin{equation}
         \frac{\partial n(\theta,\phi)} {\partial t}=
           \frac {1} {sin\theta}\,
           \frac  {\partial} {\partial \theta}
           \left\{ sin\theta
               \left[
                 h'
                 \frac{\partial V^{mag}}{\partial \theta}
                 n
             + h'\frac{k_BT}{v}
               \frac{\partial n}{\partial \theta}
             \right]
          \right \}
\end{equation}

These expressions will be used in the following paragraph for
deriving the N\'eel-Brown relaxation time.

\subsubsection{The double-well potential and the relaxation times }

\begin{figure}[h!t]
         \begin{center}
         \begin{tabular}{c}
         \includegraphics[height=5cm]{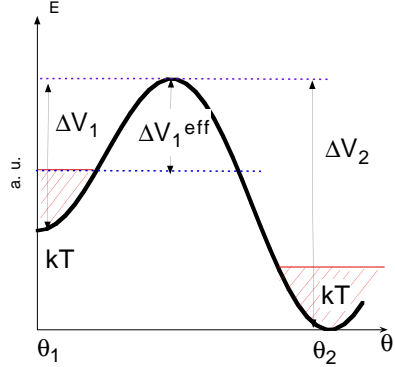}
         \end{tabular}
         \end{center}
         \caption[Double-well potential with stochastic fluctuations]
{ \label{fig:Potentiel} Double well potential (continuous line) with
stochastic fluctuations (dashed area), and the definition of the
barrier heights.}
         \end{figure}

The double well potential (see fig. \ref{fig:Potentiel}) is the
first approximation of the ferromagnetic particle energy
$V^{mag}(\theta,\phi)$ beyond the harmonic potential, but it is also
a realistic magnetic potential in the case of an uniform
magnetization with uniaxial anisotropy \cite{SPIE}:

\begin{equation}
V^{mag} = K \sin^{2}\theta- M_sH_{ap}\cos(\theta-\varphi)
\label{PotMag}
\end{equation}

where
$\theta$ is the direction of the magnetization, $\varphi$ the
direction of the applied field $H_{ap}$, and K is the anisotropy
constant in energy per unit volume.

\smallskip

In order to apply Brown's method, we consider a potential
which has minima at $\theta_1$ and $\theta_2=\pi-\theta_1$, and a
maximum at $\theta_m$. Following Kramers transition theory, Brown
\cite{Brown} assumes that most of the representative points on the
unit sphere are concentrated at the energy minima of
$V^{mag}(\theta)$ where they are in thermal equilibrium so that
locally $n$ takes the form of the Maxwell-Boltzmann distribution.
Thus only a minute fraction of the representative points is outside
the energy minima allowing a small diffusion current between them so
manifesting the non-equilibrium conditions.
\begin{equation}
       n(\theta,\phi)=
       \left\{
       \begin{aligned}
         n(\theta_1)e^{-\frac{v}{k_BT}[V(\theta)-V(\theta_1)]}
              &\mathrm{,\ \ for\
}\theta\in(\theta_1-\epsilon,\theta_1+\epsilon)\\
         n(\theta_2)e^{-\frac{v}{k_BT}[V(\theta)-V(\theta_2)]}
              &\mathrm{,\ \ for\
}\theta\in(\theta_2-\epsilon,\theta_2+\epsilon)\\
       \end{aligned}
       \right.
\end{equation}

The number of particles $N_1 \leftrightarrow N_2$ from the first well,
respectively the second is:
\begin{equation}
      \left \{
      \begin{aligned}
       N_1=2\pi n(\theta_1)e^{\frac{v}{k_BT}V(\theta_1)}I_1,
         \mathrm{\ \ \ where \ \ \ }
         I_1=\int_{\theta_1-\epsilon}^{\theta_1+\epsilon}
                e^{-\frac{v}{k_BT}V(\theta)}sin\theta d\theta\\
       N_2=2\pi n(\theta_2)e^{\frac{v}{k_BT}V(\theta_2)}I_2,
         \mathrm{\ \ \ where \ \ \ }
         I_2=\int_{\theta_2-\epsilon}^{\theta_2+\epsilon}
                e^{-\frac{v}{k_BT}V(\theta)}sin\theta d\theta
      \end{aligned}
\right.
\end{equation}

Assuming that the flow between the two minima $\theta_1$ and
$\theta_2$ is \textit{quasistationary}, approximated by a
divergenceless current\cite{Brown}, the total current of particles
over the potential barrier can be written:
\begin{equation}
       I=2\pi \sin\theta J_{n}^{\theta}
\end{equation}
Rewriting the first equation from \ref{fluxanisotropy}, one obtains:
\begin{equation}
          \frac{\partial n}{\partial \theta} +
          \frac{v}{kT} \frac{\partial V^{mag}}
                            {\partial \theta}\, n
        = -\frac{Iv}{2 \pi h'k_BT  sin\theta}
        \label{QuasiStat}
\end{equation}
which defines the \textbf{activation regime}. Introducing $I_m$ as

\begin{equation}
      I_m=\int_{\theta_1+\Delta\epsilon}^{\theta_2-\Delta\epsilon}
                 \frac{ e^{ {\frac{v}{k_BT}V(\theta)} }  }
                      {sin\theta}
                 \, d\theta
\end{equation}
Eq. (\ref{QuasiStat}) yields

\begin{equation}
       I=-\dot{N_1}=\dot{N_2}=
          -\frac{h'k_BT}{vI_m}
          \left(
            \frac{N_2}{I_2}-
            \frac{N_1}{I_1}
          \right)
\end{equation}
which has the form
\begin{equation}
       \dot{N_1}=-\dot{N_2}=\frac{N_2}{\tau_2}-\frac{N_1}{\tau_1}
\end{equation}
with
\begin{equation}
      \left\{
       \begin{aligned}
         \tau_1&=\frac{I_1I_mv}{h'k_BT}\\
         \tau_2&=\frac{I_2I_mv}{h'k_BT}
       \end{aligned}
      \right.
\end{equation}

Because of the rapid decrease of the exponential factor with
distance from the minima of $V^{mag}$, we may in $I_1,\,I_2,\,I_m$
replace $V^{mag}(\theta)$ by its Taylors's series about
$\theta_1,\theta_2$, respectively $\theta_m$ truncated at the
$\theta_{2}$ term, and replace the upper limit of the integrals by
$\infty$. With these approximations , we find

\begin{equation}
\left\{
\begin{aligned}
       \tau_1&= \tau_{01} \,
              e^{\frac{v(V(\theta_m)-V(\theta_1))}{k_BT}}\\
       \tau_2&=\tau_{02} \,
              e^{\frac{v(V(\theta_m)-V(\theta_2))}{k_BT}}
\end{aligned}
\right. \label{relaxtimes}
\end{equation}
where the waiting times are given by the expressions:
\begin{equation}
\left\{
\begin{aligned}
       \tau_{01}&=\frac{2\pi}{h'}
               \left[
                -V"(\theta_1)V"(\theta_m)
              \right]^{-\frac{1}{2}}
              \frac{\sin\theta_1}{\sin\theta_m} \\
       \tau_{02} &=\frac{2\pi}{h'}
               \left[
                -V"(\theta_2)V"(\theta_m)
              \right]^{-\frac{1}{2}}
              \frac{\sin\theta_2}{\sin\theta_m}
\end{aligned}
\right. \label{waitingtime}
\end{equation}

Equation Eq. (\ref{relaxtimes}) is a formula for a \textit{symmetric
      \ bistable} potential which has minima in $\theta_1$,
$\theta_2=\pi-\theta_1$ and a maximum in $\theta_m=\pi/2$.
\smallskip

For $\varphi\neq0$, the potential $V^{mag}(\theta,0)$ has an
asymmetric bistable form, and all arguments leading to
Eq. (\ref{relaxtimes}) also apply for an arbitrary $\varphi$
\cite{Coffey}. The general equations are very similar to Eq.
(\ref{relaxtimes}), the only difference in the analytic expression is
that instead of the symmetric angles $\theta_1$, $\theta_2$, we have
the asymmetric angles $\theta_A$, respectively $\theta_B$.

It has to be emphasized that $\tau_1$ and $\tau_2$ are the relaxation
times related to the first and second potential barrier (starting from
the first or the second minima), and that Eq.  (\ref{relaxtimes})
constitutes the N\'eel-Brown law for the particular case of the
\textit{symmetric bistable} well.
Interestingly, in all cases \cite{Coffey}, the Gilbert damping is
reduced to the prefactor of the exponential, and consequently plays only a
negligible role in the activation process.

In many cases, the potential is highly asymmetric, and the
N\'eel-Brown law is written in the following asymptotic form:

\begin{equation}
\tau =\tau_{0} \exp \left (\frac{\Delta
V_0(1-H/H^{0}_{sw})^{\alpha}}{kT} \right ) \label{activation}
\end{equation}

with three phenomenological parameters $\alpha \approx 3/2$, $\Delta
V_{0}$ and $H_{sw}^{0}$ \cite{WW,WW2}.  The laws (\ref{relaxtimes})
and (\ref{activation}) are well established experimentally in usual
magnetic sub-micro structures. More surprisingly, they have been
also observed in CIMS experiments under spin-injection with high
effective temperature $T_{eff}$ (2000K to 20000K) instead of 300K to
340 K \cite{Myers,SPIE,MSU,Fabian,Pufall,Guittienne} (see next section).

\begin{figure}[h!t]
         \begin{center}
         \begin{tabular}{c}
         \includegraphics[height=8cm]{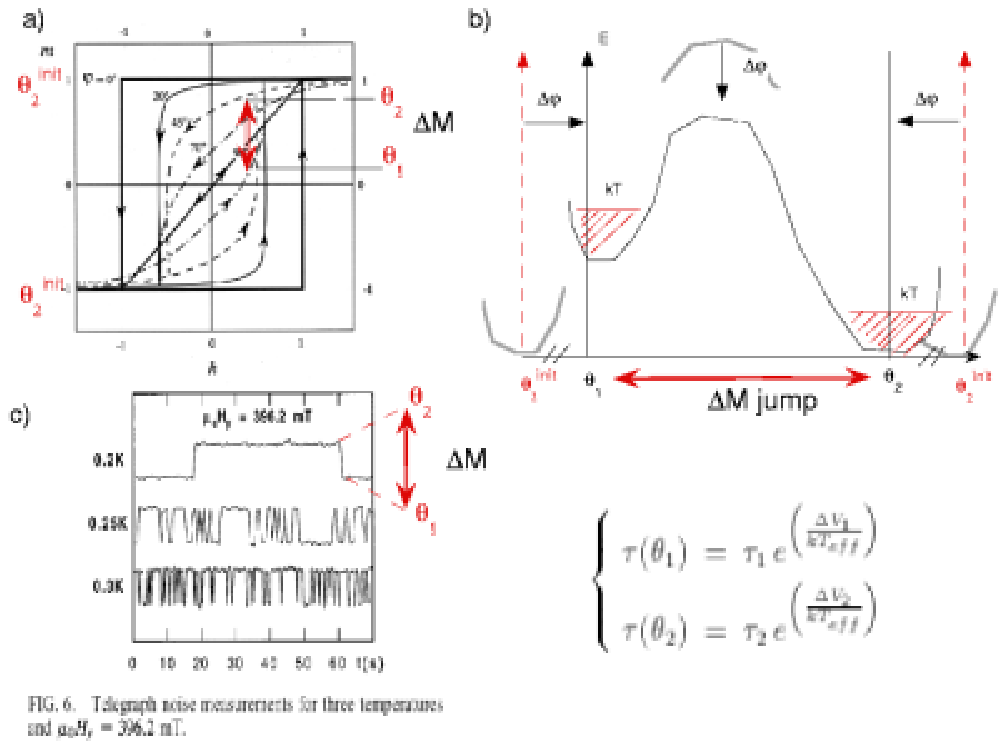}
         \end{tabular}
         \end{center}
         \caption[Slow ferromagnetic relaxation]
         {\label{fig:MagJump}
         Slow ferromagnetic relaxation measurements in uniaxial anisotropic
         single magnetic domain.  (a) hysteresis loop at different angles
         $\varphi$ of the applied magnetic field.  The saturation (i.e.
         initial states without excitation in the relaxation protocol:
         $H_{y} = 0$) corresponds to the magnetization $\theta_{1}^{init}$
         or $\theta_{2}^{init}$.  (b) double well potential after the
         application of the excitation $\Delta \varphi$: the whole
         potential landscape is modified (the barrier height and the two
         equilibrium configurations).  (c) two level fluctuations measured
         between the two configurations (the magnetization jumps from
         potential well at $\theta_{1}$ to the potential well at
         $\theta_{2}$.  Right: the N\'eel-Brown law is verify after
         performing  statistics over a significant number of jumps, in
         order to access to the mean relaxation times $\tau(\theta_{1})$
         and $\tau(\theta_{2})$.  Reprint with the permission from Ref.
         \cite{WW2} Phys.  Rev.  Lett.  78, 1791 (1997),
         Copyright@American Physical Society }
         \end{figure}

Fig.  \ref{fig:MagJump} describes how to measure the ferromagnetic
potential landscape with the help of slow magnetic relaxation
measurements (the so called magnetic after-effect measurements),
{\it performed on a single magnetic domain nanostructure}
\cite{SPIE}.  In the example shown, the angle and amplitude of the
magnetic field is set in order to obtained the two-level fluctuation
effect (measurements reported in Wernsdorfer et al.  \cite{WW2}).
The principle of the measurements is sketched in Fig.
\ref{fig:MagJump} for a sample with uniaxial anisotropy : the
hysteresis loop describes the succession of equilibrium
magnetization configurations (or quasi-static states) as a function
of the magnetic field (the field is normalized to the anisotropy)
for different angles $\varphi$ of the applied field.  The angle
$\theta$ describes the direction of the magnetization.  For a given
angle of the field $\varphi$, the hysteresis is composed of the
reversible configurations and two symmetric irreversible jumps over
the potential barrier.  The jump occurs from one equilibrium
(metastable) state defined by the angle $\theta_{1}$ to the other,
defined by the angle $\theta_{2}$ (Fig. \ref{fig:MagJump}(b)).  At
zero Kelvin (without fluctuation), this angle is related to the
applied field through a relation that depends on the switching mode
(the switching mode concerns the magnetic configurations occurring
during the jump, i.e. the modes with a typical lifetime of fraction
of nanoseconds \cite{Brown,Aharoni}).  In contrast, the amplitude of
the jump is easily measured (it is given by the difference $M_{s}
(cos \theta_{2} - cos\theta_{1})$), and gives direct access, from
the hysteresis loop, to the quasi-static configurations (i.e. to the
position of the potential wells). Changing the field (amplitude and
angle) necessarily changes the equilibrium position (defined by
$\nabla V(H) =0$) and the angles $\theta_{1}$ and $\theta_{2}$. Note
that the case of uniaxial symmetry with $\varphi = 0$ is the unique
pathological case, where the change in the {\it amplitude} of the
field does not change the equilibrium magnetic configurations.
However, since this pathological case is unfortunately that used in
most calculations for the sake of simplicity (this was also the case
here for the derivation of the N\'eel-Brown's law), the fact the
initial and final states are necessarily modified is not mentioned
in many reports of ferromagnetic relaxation experiments. Inversely,
a change observed in the two equilibrium configurations implies a
change of the potential landscape, i.e. a change of the effective
field.

The possibility of distinguishing between the effect of an effective
field and the action of the environment will be of fundamental
importance with regard to the study of magnetization reversal due to
spin injection (that are performed on single domain magnetic
nanostructures), because the main problem is to identify the typical
relaxation times that govern the mechanisms responsible for the
magnetization reversal (effective fields vs.  stochastic
fluctuations).  It is worth pointing out that it is not relevant (and
sometimes misleading) to measure relaxation effects in order to access
the valley of the potential (i.e.  the reversible magnetic configurations
induced by the action of an effective field) because the position of
the valley is given by the quasi-static measurements (the hysteresis
loop).  Furthermore, the stochastic fluctuations do not perturb the
quasi-static configurations (e.g. reversible part of the hysteresis),
except at the two critical points where the potential barrier is of
the same order as the thermal energy.  In other terms, only the position of
the irreversible jump is modified by the activation (or by the
observation time window, or equivalently by the velocity of the
sweeping field \cite{Kurkijarvi}), but not the rest of the hysteresis
loop.  Of course, this is no longer the case for larger samples,
beyond nanoscopic dimensions, because the sample is a distribution of
single domain sub-systems with a distribution of anisotropies, defects,
etc.  In contrast, the barrier height - or the amplitude of the
thermal fluctuations - can be measured only by the activation
process, through statistical measurements.  If the barrier height is
above the energy of the lattice $kT$, the two level fluctuations occur
within a typical time window that can be tuned with the amplitude of
the magnetic field or the temperature.  In order to access to the
relaxation times, statistics should be performed over a significant
number of shots.  If the exponential relaxation is verified, the
relaxation times can then be extracted.  The ratio of the relaxation
times $\tau(\theta_{1})/\tau(\theta_{2})$ gives the asymmetry of the
double well, and each relaxation time gives the corresponding barrier
height.  The N\'eel-Brown law is tested by varying the temperature
and the magnetic field (Fig.  \ref{fig:MagJump}).  The whole potential
can be rebuilt form these measurements.

\section{Spin-injection induced ferromagnetic Brownian motion}

Let us now consider a system composed of a ferromagnet in which
spin-injection (with high current density) is performed.  In such
spin-transfer experiments, the magnetization configurations are
measured while injecting the current.  Usually the magnetic
configurations are measured with GMR or AMR properties for
convenience, but micro-magnetometry or magneto-optics measurements are
also possible \cite{Siegmann}.  The magnetization is a macroscopic (thought
nanoscopic) variable described in the previous section (Sec.  III),
and the spin-injection is described by the two -or four- channel model
presented in section II. We start with the assumption that the
magnetic system is an open system, composed on one hand of the
spin-accumulated charge carriers, and on the other hand by the
ferromagnetic order parameter.  It is worth pointing out that in this
picture, we are not dealing with a theory of itinerant ferromagnets,
in which spin injection is performed.  Such a theory would allow the
dynamics of the macroscopic variable $\vec M$ to be derived from the
spin-dependent electronic populations $\delta n$ (e.g. a theory in
which the LLG equation would be derived from the Hamiltonian of the
electronic system \cite{Kambersky}).  Such a theory would be very
difficult because statistical projections
   should necessarily be performed in
order to reduce the microscopic degrees of freedom to fluctuations and
damping for the macroscopic order parameter
\cite{Forster,Fick,Haenggi,Balian}.  The simple
phenomenological model proposed here can be viewed as a first step in
this direction, by defining the different sub-systems, that are
coupled through relaxation processes (see first Figure in Sec.
I).

\subsection{Current-dependent effective field?}
\label{subsec:EffectField}

Before dealing with the dissipative coupling invoked above, a first
trivial mechanism that couples the two sub-systems should be
discussed.  This is a deterministic coupling through a common
electro-magneto-chemical potential $\mu$, from which the associated
effective fields are derived $ \partial_{x} \mu_{0}$, $\partial_{x}
\Delta \mu $ (electric) $\partial_{\theta} \mu$ (magnetic), $ \Delta
\mu$ (spin pumping).  This concept follows that introduced by
Guggenheim \cite{Guggenheim} when he defined the electro-chemical
potential in order to describe electrochemical or electrophoretic
processes.  For instance, the local electric
field $\partial_{x} \Delta \mu$ is spin-dependent, and might have an
action on the magnetic order parameter.  On the other hand, the
magnetic field $\partial_{\vec u} V^{mag}$ is able to have an action
on the spin polarized current.

      These effects are defined by the deterministic part $\mu$ of the
      chemical potential $\tilde \mu$ and can hence be reduced to the
      action of the effective field $\vec H_{eff}(I) = -\vec \nabla \mu$
(drift part
      of the Fokker-Planck equation, to be opposed to the diffusion part),
      according to Eq.  (\ref{FPEnous}).  In other terms, within this
      hypothesis, CIMS experiments reproduce the slow magnetic relaxation
      (or magnetic after-effect) experiments described in the last
      paragraph of the previous section, where the magnetic field
      excitation is replaced by the injection of current.

      In the context of thermokinetics, the two mechanisms are
      identical: the current or the field drives the magnetization from
      a stable initial position ($\theta^{init}$) to the metastable
      state $\theta_{1}$ or $\theta_{2}$ by modifying the position of
      the potential landscape.  If a relaxation is observed, this is due
      to the usual activation process only, i.e. the jump over the
      barrier induced by the thermal fluctuation (helped by the Joule
      heating due to the current) \cite{Li,Apalkov,Serpico}.
      Accordingly, as underlined at the end of the previous section,
      this deterministic mechanism due to current injection should first
      be observed on the stable magnetic configurations, by measuring
      the modification of the potential wells (e.g. measuring the
      modification of the hysteresis loop of a single domain nanolayer)
      under current injection.

\begin{figure}[h!t]
       \begin{center}
        \begin{tabular}{c}
          \includegraphics[height=8cm]{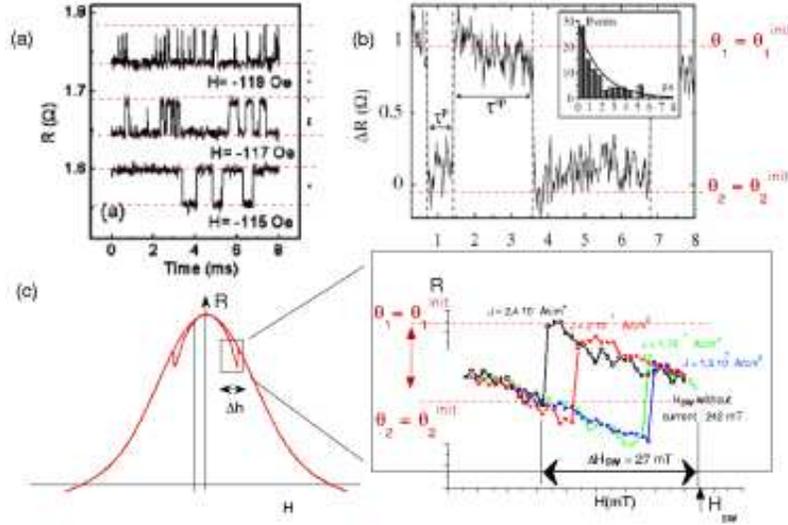}
        \end{tabular}
       \end{center}
       \caption[Current induced ferromagnetic relaxation]
         {\label{TLF} Ferromagnetic relaxation measurements under
spin-injection.  (a)
         Two level fluctuation measured with GMR in a trilayer system
         (Permaloy/Cu/Permaloy, from \cite{MSU}).  (b) Two level
         fluctuations measured with GMR in a trilayer structure (Co/Cu/Co,
         from \cite{Fabian}).  (c) Hysteresis loop measured with AMR (left)
         and zoom around the irreversible jump measured under
         spin-injection (6 microseconds pulse per point) (details reported
         in \cite{Guittienne}).  The hysteresis measured with AMR shows the
         succession of the equilibrium configurations.  In all cases, the
         initial and final states (the equilibrium states) are that
         measured without current injection ($\theta = \theta^{init}$).
         Equilibrium configurations are not modified by the spin-injection.
         Reprint with the permission from Ref.  \cite{MSU} S. Uhrazdhin, et
         al.  Phys.  Rev.  Lett.  91 146803 (2003) and
         A. Fabian, et al. Phys. Rev. Lett. 91 257209 (2003),
Copyright@American
         Physical Society.}
         \end{figure}

      However, it is worth pointing out that the modification of the
      potential landscape is not observed: the basic prediction of the
      determinist action of the spin-injection has not been verified until
      now \cite{Albert,Myers,Sun,MSU,SPIE,Fabian,Guittienne}.  Fig.
      \ref{TLF}
      shows two shots with two level fluctuations in a time interval of
      milliseconds (a) (\cite{MSU}) and microseconds (b) (\cite{Fabian}),
      obtained by different groups with GMR measurements on trilayer
      nanopillars, and single irreversible jumps measured in an hysteresis
      loop of a Ni nanowires (c) (\cite{Guittienne}): each point in the
      zoom (right part of Fig. \ref{TLF} (c)) is a 6 microseconds pulse.  The
      hysteresis is performed with current varying from 2.4 $10^{7}$
      A/cm$^{2}$ to 1.5 $10^{7}$ A/cm$^{2}$ (the position of the jump
      without current injected is shown by the arrow).  In all cases, the
      metastable states (described by the angles $\theta_{1}$ and
      $\theta_{2}$ defined in the previous section) coincides with
      $\theta_{1}^{init}$ and $\theta_{2}^{init}$ measured before the
      application of the current. All happens as if the potential
      landscape were not modified!

\smallskip

In contrast, the action of the current is huge (some fraction of an eV to
few eV, i.e. from 25 \% to more than 100 \% of the anisotropy energy 
of the ferromagnet), and is observed only for the irreversible jump.  By acting on
the jump and not on the potential landscape, the effect of the current
mimics the action of a temperature. 

The arguments developped above are based on an analysis performed with
neglecting the precession terms.  If we assume an hypothetical
precession maintained in a stationary regime, the magnetization is
then driven by the precession, and the trace in Fig.  6 should be
interpreted as trajectories in the phase space.  Within this context,
the precession induces {\it intermittency}, i.e. a highly specific
chaotic behavior in which the time spent out of the attractors is
negligible.  Such an approach is discussed in some theoretical works
about CIMS. This interpretation is however evacuated here because the
measured traces (Fig.  6) mimic exactly the full stochastic process
composed of the two-level fluctuation (that follows the N\'eel-Brown
activation law), {\it superimposed to the same white noise for both
quasi-static states} \cite{Noise}.  The observation of identical noise
in both states is in contradiction with published simulations
performed with the hypothesis of spin transfer torque \cite{Zhu,Lee}.  Furthermore,
after spending years in measuring activation processes in 
ferromagnetic nanostructures,
it is difficult to accept that a deterministic behavior replaces the
activation and mimics so perfectly the full stochastic behavior.

Accordingly, we focus our
attention on the {\it dissipative, or irreversible, spin-transfer}
effect, for which the action of the current is expressed through the
diffusive terms of the Fokker-Planck equation .

\subsection{Negative damping}

      \begin{figure}[h!t]
       \begin{center}
        \begin{tabular}{c}
          \includegraphics[height=8cm]{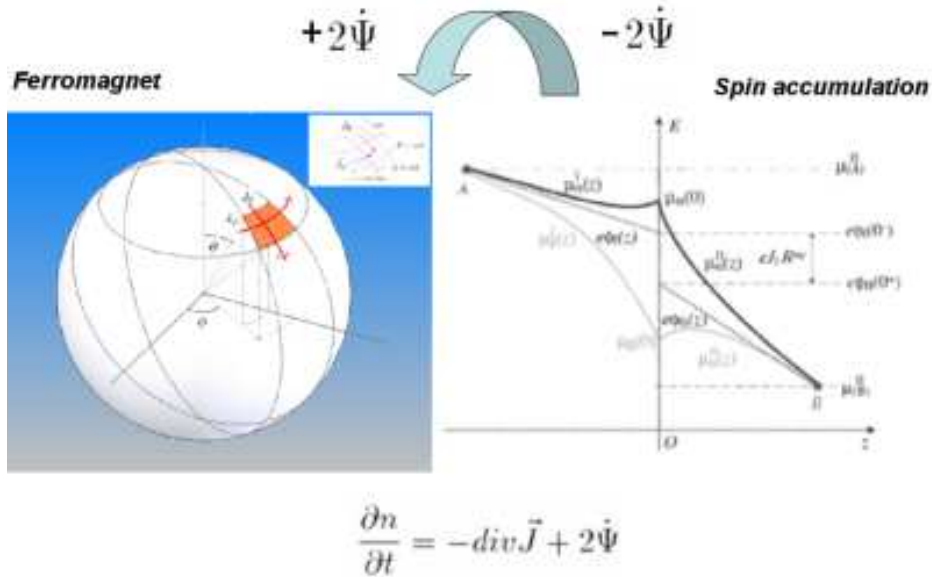}
        \end{tabular}
       \end{center}
       \caption[Dessin ST]
    {\label{ST} Under spin-injection, the ferromagnetic layer is
    an open system coupled to the spin-accumulation
    reservoir.  The variation of the density of magnetic moments
    (oriented at a given direction in the unit sphere) is given by the
    divergence of the magnetic flux added the correction due
    to the relaxation rate $\dot \Psi$ of spins relaxing from one system
    to the other
    at the interfaces.}
         \end{figure}
\smallskip

In order to describe the ferromagnetic layer coupled to the
spin-dependent electric sub-system, we rewrite the conservation law of
the ferromagnetic systems (Eq.  \ref{ContMag}) with adding the
relaxation term $\dot \Psi$ due to spin-transfer, while substracting
it from the spin-accumulation's conservation law:

\begin{gather}
      \left \{
      \begin{aligned}
\frac{\partial n^{ferro}}{\partial t} =& -div(\vec{J_{n}}) + \dot \Psi \\
\frac{\partial n^{elec}}{\partial t}=& - div(\vec{\delta J}) - \dot \Psi
\end{aligned}
\right .
\label{ContST}
\end{gather}

According to the first law of thermodynamics, the contribution due to
the electronic relaxation $\dot
\Psi$ should be taken into account in the entropy
production of the ferromagnet.  Inserting Eq.  (\ref{ContST}) into Eq.
(\ref{condiv}), the entropy production is:

\begin{align}
     \left\{
      \begin{aligned}
       \vec{J_{s}}=&\frac{1}{T}\,\vec{J_{e}}
            -\frac{\tilde\mu}{T}\,\vec{J_{n}}\\
       \mathcal{I}=&\,\vec{J_{e}}\cdot\vec{grad}
           \left(
             \frac{1}{T}
           \right)
        -\vec{J_{n}}\cdot \vec{grad}
          \left(
            \frac{\tilde\mu}{T}
          \right) + \frac{\Delta \mu}{T} \dot \Psi
      \end{aligned}
     \label{SProdST}
     \right.
\end{align}

where $\Delta \mu = \mu_{elec} - \tilde \mu$ ( e.g. $\Delta \mu =
\mu_{s \downarrow} -\mu_{d \downarrow}$).  The Onsager
transport equations are modified accordingly.  The flux $\vec{J_{n}}$
and $\dot \Psi$ are not of the same tensorial nature (in a first
approach): the first is a
vector defined on the sphere, and the second is a scalar.  According
to the Curie principle, Onsager transport coefficients that couple the
two processes should not exist.  The modification due to the
contribution of the electronic relaxation is taken into account by a
third relaxation term (we still assume $T$ constant) :

\begin{equation}
     \left \{
      \begin{aligned}
         J_{n}^{\theta}\,
         &=
            -L_{\theta\theta }
             \frac{\partial \tilde \mu}{\partial \theta}
            -L_{\theta \phi }
             \frac{1}{sin\theta}
             \frac{\partial \tilde \mu}{\partial \phi}
           \\
         J_{n}^{\phi} \,
         &=
            -L_{\phi \theta} \frac{\partial \tilde \mu}{\partial \theta}
            -L_{\phi \phi}
             \frac{1}{sin\theta}
             \frac{\partial \tilde \mu}{\partial \phi}
          \\
            \dot \Psi  \,
         &= L \Delta \mu
      \end{aligned}
     \right. \label{fluxST}
\end{equation}

where $\Delta \mu$ is defined as the chemical affinity of
the corresponding reaction, or equivalently, defined as the pumping
force associated to the flux  $\dot \Psi$ of spins transferred
between the spin-polarized electric system and the ferromagnet.  In
other terms, $\Delta \mu$ {\it is the force responsible for
irreversible spin-transfer}.  Integrating over the whole
ferromagnetic layer, the Fokker-Planck equation, Eq.
({\ref{EqGene}}), rewrites:

\begin{equation}
\frac{\partial n^{ferro}}{\partial t} =  \frac{\partial n_{0}}{\partial
t} + \int_{A}^{B} L_{sd} \, \Delta \mu (z) dz = 0
\label{ContEff}
\end{equation}

where  the first term in the right hand side $\frac{\partial n_{0}}{\partial
t}$ leads to the standard
rotational Fokker-Planck equation (defined by the second equation in
Eqs. (\ref{rotFPE})), and the second term is the contribution coming
from the electronic relaxation.  {\it This equation is the main
result that defines the irreversible spin-transfer effect for an open
system}. Eq. (
\ref{ContEff}) will not be solved here.  The aim of the following
developments is to define negative damping and effective
temperature.

The dynamic equation is obtained in the same way as in the
previous section, by writing the flux of representative points on
the unit sphere. A term $\frac{d \vec{u}^{elec}}{dt}$ (not explicit
here) related to
the spin-polarized electronic contribution is added to the Eq.
(\ref{LLG}) and leads to the following generalized LLG equation:

\begin{equation}
         \frac{d \vec{u}}{d t}= \frac{\partial
         \vec{u}^{elec}}{\partial t}
             -g' \vec{u} \times
                \left(
                  \vec \nabla V^{mag}
                \right)
         +h'\vec{u} \times
           \left[
             \vec{u} \times
             \left(
               \vec \nabla V^{mag}
             \right)
           \right]
          \label{GenLLG}
\end{equation}

Since the effect of the environment conserves the modulus of the
magnetization, i.e. $ \vec{u} .  d \vec{u}/ d t = 0$ \cite{Callen},
the contribution
of $\frac{d \vec{u}}{d t}^{elec}$ reduces to the two damping factors,
parallel to $\vec{u} \times \nabla V^{mag}$ and parallel to $\vec{u}
\times (\vec{u} \times \vec \nabla V^{mag})$, and a stochastic force
$f(t)$.  This necessarily leads to the introduction of two parameters
$\alpha_{1}$ and $\alpha_{2}$ such that :

\begin{equation}
        \frac{d \vec{u}}{d  t}=
             -\left (g'+\alpha_{1} \right ) \vec{u} \times
                \left(
                  \vec \nabla V^{mag}
                \right)
         + \left ( h' + \alpha_{2} \right ) \vec{u}\times
           \left[
             \vec{u}\times
             \left(
               \vec \nabla V^{mag}
             \right)
           \right] + \vec{f}(t)
          \label{ReducLLG}
\end{equation}

where the coefficients $\alpha_{1}$ and $\alpha_{2} $ can be thought
of as {\it negative damping} or {\it positive damping}, depending
wether the spin transfer $\dot \Psi_{sd}$, integrated over the whole
layer with the two junctions, is transferred from the electric
system to the magnetic system (negative damping) or inversely
(positive damping).  In other words, it depends on the balance of
spin accumulation at the two interfaces of the ferromagnetic layer.
Note that as far as the damping coefficients are not explicitly
defined, the validity of the argument used above (the Callen's
argument \cite{Callen}) is not restricted to the relaxation and spin
accumulation mechanisms described in the first sections of this
work, but is much more general.  In particular, the equation is
formally similar to that described in the framework of the exchange
torque or spin torque theory
\cite{Sloncz,Bazaliy,Waintal,Levy2,Stiles,Bauer,Polianski}. However,
beyond the vectorial argument proposed above, the proper derivation
of the reduction from Eq.  (\ref{GenLLG}) to Eq. (\ref{ReducLLG})
(see e.g. \cite{Rubi2}) is still to be performed.

       \subsection{Effective thermostat}

The N\'eel-Brown activation laws describe out-of-equilibrium spin
systems ($\partial \tilde \mu \ne 0$), and are valid for high enough
potential barriers $kT \ll v V^{mag} $ , or long time scales $ \Delta
t / \tau_{0} \gg 1$, where $\Delta t$ is the measurement time window,
and $\tau_{0}$ is the relaxation time scale that describes the
coupling to the lattice.  If the volume $v$ tends to zero, the energy
barrier decreases down to a value such that $v V^{mag} \le kT ln
(\Delta t/\tau_{0})$, and the magnetization is at $"\it equilibrium"$
with the lattice for the measurement time window $\Delta t$.  The
system is no longer metastable but {\it superparamagnetic}.  The
equilibrium imposes the condition $ \partial \tilde \mu = 0$, and the
statistical distribution is the Maxwell-Boltzmann distribution $
n=N_{t} \exp \left (-v V^{mag}/kT \right )$.  The magnetization
behaves like a paramagnetic spin, with the ferromagnetic order
parameter $\vec M$ instead of the spin $\vec s$.  In the case of 3d
metallic ferromagnetic nanostructures (Co, Ni, Fe \ldots) of sizes
typically around 10 nm radius at room temperature, the system is
superparamagnetic for time scales of magnetometric
measurements (above $10^{-5}$ sec.).  The system is nevertheless
ferromagnetic and follows the N\'eel-Brown laws if the measurements
are performed in a shorter time window $\Delta t$ (from micro-seconds
to nanoseconds in the case of $10 nm^{3}$ particles invoked above).

However, if the relevant time window is shorter than the typical
ferromagnetic relaxation time scale $\tau_{0} $ ($\Delta t \le
\tau_{0}$), the precession now governs the quasi-ballistic dynamics,
which is qualitatively different (because it is not driven by the
fluctuations).  The collective modes measured are that observed with
ferromagnetic resonance; e.g. dynamics of thermal spin waves are
observed in GMR structures
\cite{Bertram,Smith,Chappert,ChappertBal,Russek}.  It is no longer
activated, whatever the potential barrier and the volume
$v$, and the thermalization process vanishes at short time scales
("quasiballistic magnetization reversal" regime \cite{ChappertBal}).
The temperature of the system (if any \cite{Casas}) is not
necessarily the
temperature of the lattice $T_{eff} \ne T$: the system is decoupled
from the heat bath.  A similar situation justified the introduction of
the concept of spin temperature $T_{s}$ in the early 50's with the
first spin resonance experiments \cite{Purcell}.  Note that if the
system is also isolated from other sub-systems at comparable time
scales, has an upper bound, and if the populations (up and down) can be
inverted (like in nuclear spin systems), the spin temperature of the
system can even be negative \cite{Ramsey,Abragam} (but the spin
temperature is usually positive and higher than the lattice
temperature \cite{Abragam}).

In the situation of interest, with spin-polarized currents in 3d
metallic nanostructures, the ferromagnetic order parameter is
coupled to the lattice through the polarization of the electronic
degrees of freedom \cite{Kambersky,Kambersky2,Shul}.  Without being
coupled to the magnetization (e.g. in the non- magnetic side of a
junction), the spin-accumulation sub-system relaxes toward
equilibrium with the relaxation time $\tau_{sf}$ of some picoseconds
(as described in the first section with the two channel model).
This relaxation time is shorter than the thermalization of the
magnetization with the lattice $\tau_{0}$ (nanoseconds).  On the
other hand, the coupling with the ferromagnetic sub-systems
corresponds to a relaxation time $\tau_{sd}$ (comprised between the
electronic relaxation time $\tau_{e}$ and $\tau_{sf}$) shorter than
$\tau_{0}$ : as a consequence, this relaxation "thermalize" the
ferromagnetic order parameter with the spin-accumulation sub-system,
that takes the role of the heat bath in place of the lattice.  This
picture is that schematized in Fig \ref{model} in the first section.

In the activation regime, it is possible to assume that the
spin-accumulation sub-system is a reservoir of energy, and that the
ferromagnetic order parameter is thermalized with it (see Fig.
\ref{model}). Providing that the spin-accumulation sub-system is not
thermalized with the lattice, {\it the zeroth law of thermodynamics
is not valid} \cite{Casas}, and it is possible to identify it as a
thermostat at temperature $T_{eff}$ in equilibrium with the
ferromagnetic system.

The equilibrium condition imposes that $\mathcal I = 0$
\cite{Vilar}. The entropy production $\mathcal I$ of the
spin-dependent electric sub-system was calculated in Sec. II.  With
the temperature $T_{eff}$ corresponding to the effective equilibrium
temperature, the chemical potential writes \cite{Mazur} $ \ \ \ \
\Delta \tilde \mu^{eff} = \Delta \mu^{eff} + kT_{eff}
ln(n_{\alpha}/n_{\gamma})$.

\begin{equation}
          T_{eff}. \mathcal I =
          \left ( - \frac{\partial \Delta \tilde \mu^{eff}}{\partial z} +
          2 \epsilon \Delta \tilde \mu^{eff} \right) \delta J
            + \Delta \tilde  \mu^{eff} \dot \Psi = 0
\end{equation}

where the Joule heating contribution $- \frac{\partial
\mu_{0}}{\partial z} J_{0}$ has been removed because it does not
contribute to the magnetic system and is coupled to the lattice (the
whole analysis should also include the
Peltier effects: energy can also be extracted from the lattice to
the magnetic system).

The entropy production vanishes for the following sufficient condition:
\begin{equation}
\Delta \tilde \mu^{eff} = 0 \label{TeffCond}
\end{equation}
The condition Eq. (\ref{TeffCond}) leads to the expression of the
equilibrium temperature $T_{eff}$:
\begin{equation}
k T_{eff}= - \frac{\Delta \mu}{ln(n_{\alpha}/n_{\gamma})} \approx -
2 \Delta \mu \frac{n_{0}}{\delta n} \label{TeffGen}
\end{equation}

where the inversion of population implies that $\delta n \le 0$.  This
equation is simply the equilibrium Curie-Weiss law that accounts for
the paramagnetic behavior of the spin-accumulation $g \mu_{B} \,
\delta n$ i.e. the first order approximation of the averaging over the
Boltzmann distribution at temperature $T_{eff}$ ( $\delta n$ is the
s-d spin accumulation that would be measured with a lattice
temperature $T_{eff}$).  The evaluation of $\delta n $ would
necessitate the non-equilibrium distribution at temperature T to be
calculated.  This task is beyond the scope of the present review.
However, according to the evaluation performed below, an energy
$kT_{eff } \approx 1$eV can be expected by calculating the
ferromagnetic energy under a current of 1 mA due to the spin transfer
in the internal field of $1 T$ of the ferromagnet.

\begin{figure}[h!t]
         \begin{center}
         \begin{tabular}{c}
         \includegraphics[height=10cm]{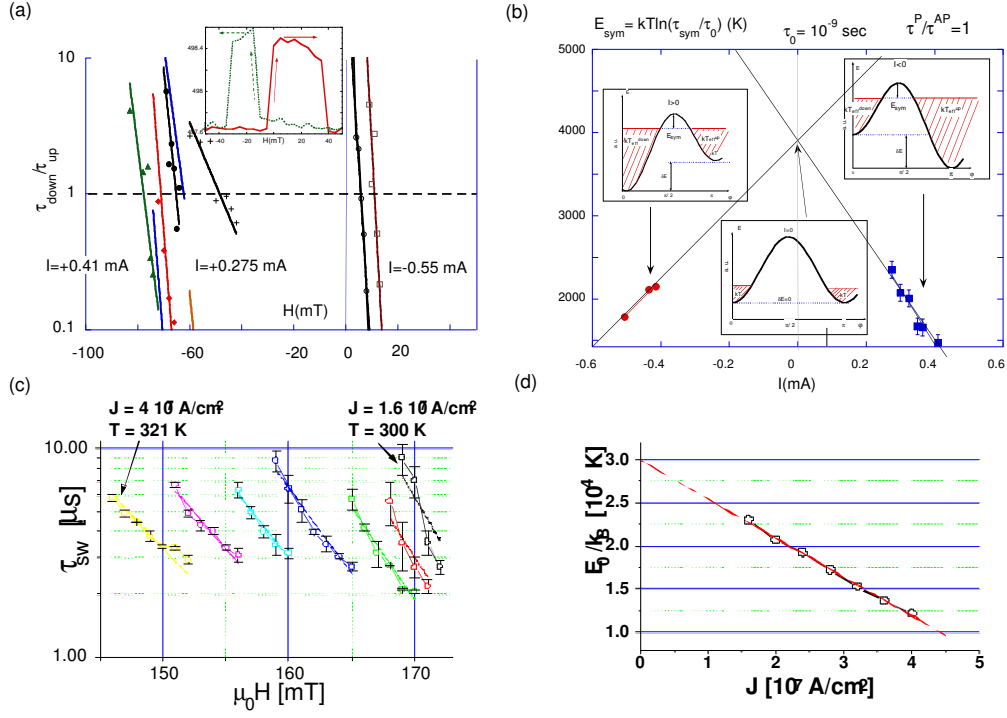}
         \end{tabular}
         \end{center}
         \caption[N\'eel-Brown activation due to current injection]
         {\label{NB} Observation of the N\'eel-Brown activation due
         to current
         injection in a Co/Cu/Co trilayers (a),(b) and in a Ni nanowire
         (c),(d).  (a) Ratio of the two relaxation times (TLF) and (c)
         relaxation time as a function of the applied field for different
         currents fitted with the N\'eel-Brown formula and
         effective barrier height as fitting parameter.  (b) and (d)
         Variation of the effective barrier height in Kelvin.  In the case
         of the two level fluctuations, the barrier is measured for the
         symmetric relaxation times (b) (i.e. for different applied
         fields).  The effect of the fluctuations are sketched in the
         insets (dashed lines).  Reprint with permission from Ref.
         \cite{SPIE} J.-E. Wegrowe Phys.  Rev.  B 68, 214414 (2003) and
         Ph.  Guittienne et al., IEEE Trans.  Mag-37, 2126 (2001),
         Copyright@American Physical Society.}
         \end{figure}

A fundamental consequence of the existence of the effective
temperature is that the solution of the stochastic equation of the
magnetization is known, and is given by the standard
activation equation Eq.  (\ref{QuasiStat}) with the effective
temperature $T_{eff}$ instead of the lattice temperature $T $:

\begin{equation}
\frac{\partial N}{\partial \theta} + \frac{1}{kT_{eff}}
\frac{\partial V}{\partial \theta} N = \frac{I_{eff}}{2 k'_{eff} \pi
sin(\theta)} \label{QuasiStat3}
\end{equation}

where $I_{eff}$ is calculated with the Boltzmann distribution with the
ferromagnetic energy $exp
\left (V(\theta)/kT_{eff} \right )$.  Assuming an approximatively
constant effective temperature $T_{eff}(\theta) \approx T_{eff}$,
the equation is formally identical to Eq.~(\ref{QuasiStat}) so that
the N\'eel-Brown activation formula is recovered with $T_{eff}$
instead of T:

\begin{eqnarray}
\left\{\begin{array}{lll} \tau(\theta_{1}) &=& \tau_{01} \,
e^{\left (
\frac{\Delta V_{1}}{kT_{eff}}\right )} \\

\tau(\theta_{2}) &=& \tau_{02} \, e^{\left ( \frac{\Delta
V_{2}}{kT_{eff}}\right )} \label{activationTLFeffect}
\end{array}\right.
\end{eqnarray}
where $\tau_{01}$ and $\tau_{02}$ contain the $k'_{eff}$ dependence.
This behavior is experimentally observed
\cite{SPIE,Myers,MSU,Fabian,Pufall}.  Fig.  \ref{NB} shows the
typical N\'eel-Brown activation observed with the sample shown in
Fig. \ref{TLF} (b) and (c). The fit of the mean relaxation time as a
function of the applied field and the current amplitude injected in
the device is performed with the N\'eel-Brown formula with the effective
barrier height as fitting parameter (Fig. \ref{NB} (a) and (c)). The
effective barrier height as a function of the current is presented
in Fig. \ref{NB} (c) and (d).

Assuming that the mechanism responsible for $T_{eff}$ is the
spin-accumulation occurring at the interface composed of antialigned
ferromagnets, the voltage drop due to spin-accumulation is
approximatively equal to $\Delta \mu$ (see Eq.~(\ref{Result})):

\begin{equation}
k T_{eff} \approx  -\frac{\Delta
R^{sa}I}{ln(n_{\alpha}/n_{\gamma})} \propto  \frac{\Delta
R^{sa}I}{2}
\end{equation}

with $-ln(n_{\alpha}/n_{\gamma}) \le 1$ The proportionality between
$kT_{eff}$ and $R^{sa}$ was observed by different groups in recent
experimental investigations \cite{MSU2,Jiang} (with DC measurements,
the parameter is the "critical current" $I_{c} \propto T_{eff}^{-1}$ as
shown in Fig. \ref{NB} (d)).

How to estimate the magnetic energy of the spin-accumulation system in
usual experimental situations, where a current density of some few mA
is injected in the nanostructure
\cite{EPL,Albert,Myers,Sun,MSU,SPIE,Fabian,Pufall,Guittienne}?  This
current corresponds to some $ 10^{16}$ spin per seconds flowing
through the interface.  If one assumes that 20 \% (polarization of the
current) of the spins are maintained out-of-equilibrium within a
typical relaxation time $\tau_{sf}$ of $10^{-11} sec$, we are left
with about $2.10^{4}$ spins that define the magnetization of the
spin-accumulation sub-system in the volume defined by the
corresponding diffusion length.  An effect of the electric spin
relaxation on the ferromagnetic order parameter should consequently be
expected for a nanostructured ferromagnetic system that is only ten
to hundred times larger.  In an internal
field of $ H_{int} = 1$T, this energy $E= 10^{4} \mu_{B} H_{int}$ is
of the order 1 eV (beyond the Curie temperature) in accordance with
activation experiments performed on various systems
\cite{SPIE,Myers,MSU,Fabian,Pufall,Guittienne}.  Without current
injection, the magnetic order parameter is at room temperature and
consequently, the hot sub-system is the spin-accumulation system.

\section{Conclusion}

      An
unified thermokinetics approach of both spin-dependent charge
carriers and ferromagnetic Brownian motion has been presented in the
context of open systems.  The
spin-dependent electronic relaxation is
then introduced as a source term in the conservation equation of the
magnetic moment. This leads to the description of the effect of
spin-injection induced magnetization switching, or irreversible spin-transfer
in an open ferromagnetic layer.

The description of the spin-accumulated charge carriers is based on
the two conduction channel approximation, generalized to both intra
-and inter- band relaxation.  The application of the first and second
laws of thermodynamics, together with the conservation laws, lead to
the spin-dependent transport equations.  The relevant Onsager
transport coefficients are introduced and related to the typical
electronic relaxation times.  The effect of charge conservation and
screening is also taken into account.

On the other hand, the ferromagnetic order parameter is described on
an equal footing by introducing the conservation laws and the relevant
chemical potential, with deterministic terms accounting for the
effective field, and dissipative terms accounting for the coupling to
a relevant heat bath.  The corresponding Onsager transport
coefficients are defined with the second law of thermodynamics, and
refined with the help of the Onsager reciprocal relations.  The
rotational Fokker-Planck equation, and the Landau-Lifshitz-Gilbert
(LLG) equation are then derived within the thermokinetic theory.  The
Onsager coefficients are related to the typical time scales of the
ferromagnetic relaxation ($\tau_{0}$).  In the activation regime, the
N\'eel-Brown activation law is deduced.

In the framework of this description, the generalization of both the
Fokker-Planck equation and the LLG equation with adding the
contribution of spin-accumulation is straightforward in terms of
flux of representative points in the magnetization sphere. The
negative damping appears naturally in order to describe the exchange
of spins from the electric subsystem to the magnetic sub-system,
described as a coupling to an environment.

Furthermore, the
discussion about the different relaxation times shows that the
spin-polarized current is not thermalized to the lattice in the
stationary regime, but is thermalized with the spin-accumulation
sub-system.  The argument is that on one hand the relaxation toward
equilibrium of the spin-accumulation system (described by $\tau_{sf}$:
some tens to hundreds of picoseconds) is shorter than the
thermalization of the ferromagnetic system (described by $\tau_{0}$:
nanoseconds).  And on the other hand, the coupling between the
ferromagnetic order parameter and the spin-accumulation sub-system
($\tau_{sd}$) is shorter or equal to the $\tau_{sf}$.

An effective temperature is then derived in the activation regime
through the entropy production, and leads to the
derivation of an effective N\'eel-Brown relaxation process due to
current injection, that is experimentally observed.

\section{Acknowledgement}

We are indebted to Prof. Peter Levy for the constant interest along
this work and for a critical reading of the manuscript. 

\newpage
\appendix

\section{Microscopic approach and thermokinetic coefficients}
     \markboth{\small{B. MICROSCOPIC APPROACH AND THERMOKINETIC
     COEFFICIENTS}}{}
     \addtocounter{equation}{91}
      \subsection{Relation between $L$ and the electronic relaxation times}

        Let us consider a simple interface between two metals. Far from
        the interface, the Ohm's law is recovered : the chemical
        potentials of the channels are identical and the electric
        distribution is that of equilibrium  $n_{\alpha0}$ and $n_{\gamma0}$.
        In the following, we assume that the charge
        transfer between the two channels can be described by the
        following relation:

        \begin{equation}
          f \Delta n_{\alpha}(x)  + g \Delta n_{\gamma }(x) =0
         \label{conschrg}
        \end{equation}

The case $f = g = 1$ describes the local electrical neutrality.  We have:

      \begin{gather}
        \left\{
         \begin{aligned}
           \Delta n_{\alpha}(x)&= (\mu_{ch,\alpha} - \mu^{0})
             N_{\alpha}(E_F)\\
           \Delta n_{\gamma}(x)& = (\mu_{ch,\gamma} - \mu^{0}) N_{\gamma}(E_F)
         \end{aligned}
        \right.
      \end{gather}

      where  $\mu^{0}$ is the chemical potential in the absence of charge
      transfer and $\mu_{ch,\alpha}$ and $\mu_{ch,\gamma}$
      are the purely chemical potentials of the channels (without transfer
       $\mu_{ch,\alpha}$  = $\mu_{ch,\gamma}$ = $\mu_{0}$) ;
      $N_{\alpha,\gamma}(E_{F})$ is the density of states at the Fermi level.
The separation between the electric potential and the purely chemical
potential writes:

      \begin{gather}
        \left\{
         \begin{aligned}
           \mu_{\alpha} =  \mu_{ch,\alpha}  + eV \\
           \mu_{\gamma} =  \mu_{ch,\gamma}  + eV
         \end{aligned}
        \right.
      \end{gather}

      where V is the local electric potential.

Relation (\ref{conschrg}) gives
      \begin{equation}
        (\mu_{ch,\alpha} - \mu^{0}) f N_{\alpha}(E_F)
          +  (\mu_{ch,\gamma} - \mu^{0}) g N_{\gamma}(E_F)
        = 0
      \end{equation}

Note that the expressions of $\Delta n_{\alpha,\gamma}$ comes from
the fact that at zero Kelvin:

\begin{gather}
      \left\{
       \begin{aligned}
        n_{\alpha}(\mu_{\alpha})
          &=\int N_{\alpha}(E)f(E)dE
           =\int_{-\infty}^{\mu_{ch,\alpha}}
              N_{\alpha}(E)f(E)dE\\
        n_{\gamma}(\mu_{\gamma})
          &=\int N_{\gamma}(E)f(E)dE
           =\int_{-\infty}^{\mu_{ch,\gamma}}
              N_{\gamma}(E)f(E)dE
       \end{aligned}
      \right.
\end{gather}

     From these relations we deduce:
\begin{gather}
      \left\{
       \begin{aligned}
         \Delta n_{\alpha}&= N_{\alpha}(E_F) \, \delta \mu_{ch,\alpha}\\
         \Delta n_{\gamma}&= N_{\gamma}(E_F) \, \delta \mu_{ch,\gamma}
                           =-\frac{f}{g}\Delta n_{\alpha}
       \end{aligned}
      \right.
      \Rightarrow
      \left\{
       \begin{aligned}
         \Delta n_{\alpha}&=   \frac{gN_{\alpha}(E_F)N_{\gamma}(E_F)}
                                  {fN_{\alpha}(E_F)+gN_{\gamma}(E_F)}
                               \delta \mu_{ch,\alpha} \\
         \Delta n_{\gamma}&= - \frac{fN_{\alpha}(E_F)N_{\gamma}(E_F)}
                                  {fN_{\alpha}(E_F)+gN_{\gamma}(E_F)}
                               \delta \mu_{ch,\alpha}
       \end{aligned}
      \right.
\end{gather}

where $\delta \mu_{\alpha \gamma} = \mu_{ch,\alpha \gamma} -
\mu_{0}$.

Introducing the global transfer rate $T_{\alpha\rightarrow\gamma}$ (resp.
        $T_{\gamma\rightarrow\alpha}$) of the channel $\alpha$ to
        $\gamma $ (resp. $\gamma$ to $\alpha$), the charge conservation
        between the two channels writes ($e < 0$):

\begin{gather}
      \left\{
       \begin{aligned}
         \frac{\partial n_{\alpha}(x)}  {\partial t}
           &=-\frac{1}{e} \frac{\partial J_{\alpha}(x)}  {\partial x}
            -T_{\alpha\rightarrow\gamma}(n_{\alpha}(x),n_{\gamma}(x))
            +T_{\gamma\rightarrow\alpha}(n_{\alpha}(x),n_{\gamma}(x))\\
         \frac{\partial n_{\gamma}(x)}  {\partial t}
           &=-\frac{1}{e} \frac{\partial J_{\gamma}(x)}  {\partial x}
            +T_{\alpha\rightarrow\gamma}(n_{\alpha}(x),n_{\gamma}(x))
            -T_{\gamma\rightarrow\alpha}(n_{\alpha}(x),n_{\gamma}(x))\\
       \end{aligned}
      \right.
\end{gather}

       that leads, in the stationary regime $\frac{\partial
       n_{\alpha,\gamma}(x)} {\partial t}=0$, to the following relations:

\begin{gather}
      \left\{
       \begin{aligned}
        \frac{\partial J_{\alpha}(x)}  {\partial x}
          &= -eT_{\alpha\rightarrow\gamma}(n_{\alpha}(x),n_{\gamma}(x))
            +eT_{\gamma\rightarrow\alpha}(n_{\alpha}(x),n_{\gamma}(x))\\
        \frac{\partial J_{\gamma}(x)}  {\partial x}
          &= +eT_{\alpha\rightarrow\gamma}(n_{\alpha}(x),n_{\gamma}(x))
            -eT_{\gamma\rightarrow\alpha}(n_{\alpha}(x),n_{\gamma}(x))\\
       \end{aligned}
      \right.
\end{gather}

The Taylor expansion to the leading order of the transfer rates
$T_{\alpha\rightarrow\gamma}$ and  $T_{\gamma\rightarrow\alpha}$,
around equilibrium gives:
\begin{gather}
      \left\{
       \begin{aligned}
        \frac{\partial J_{\alpha}(x)}  {\partial x}
          &= - eT_{\alpha\rightarrow\gamma}(n_{\alpha}^0 ,n_{\gamma}^0 )
             - e \frac{\partial T_{\alpha\rightarrow\gamma}}
                     {\partial n_{\alpha}}
                 \Delta n_{\alpha}
             - e \frac{\partial T_{\alpha\rightarrow\gamma}}
                     {\partial n_{\gamma}}
                 \Delta n_{\gamma}\\
             &\qquad + eT_{\gamma\rightarrow\alpha}(n_{\alpha}^0 ,n_{\gamma}^0 )
              + e \frac{\partial T_{\gamma\rightarrow\alpha}}
                     {\partial n_{\alpha}}
                 \Delta n_{\alpha}
             +e \frac{\partial T_{\gamma\rightarrow\alpha}}
                     {\partial n_{\gamma}}
                 \Delta n_{\gamma}\\
        \frac{\partial J_{\gamma}(x)}  {\partial x}
          &=  -\frac{\partial J_{\alpha}(x)}  {\partial x}
       \end{aligned}
      \right.
\end{gather}

At equilibrium, the current of each channel is conserved, so that:
\begin{equation}
       - e T_{\alpha\rightarrow\gamma}
            (n_{\alpha}^0 ,n_{\gamma}^0 )
       +e T_{\gamma\rightarrow\alpha}
            (n_{\alpha}^0 ,n_{\gamma}^0 ) = 0
\end{equation}

Defining the electronic relaxation times:

$\tau_{\alpha\rightarrow\gamma}$, $\tau_{\gamma\rightarrow\alpha}$,
such that

\begin{gather}
      \left\{
       \begin{aligned}
        \frac{1}{\tau_{\alpha\rightarrow\gamma}} & =
          \frac{\partial
                   (   T_{\alpha\rightarrow\gamma}
                     - T_{\gamma\rightarrow\alpha}
                    )
                }
               {\partial n_{\alpha}}_{(n_{\alpha}^0 ,n_{\gamma}^0 )}\\
        \frac{1}{\tau_{\gamma\rightarrow\alpha}} & =
          \frac{\partial
                 (  T_{\gamma\rightarrow\alpha}
                  - T_{\alpha\rightarrow\gamma}
                 )
               }
                     {\partial n_{\gamma}}_{(n_{\alpha}^0 ,n_{\gamma}^0 )}\\
       \end{aligned}
      \right.
\end{gather}
we have :

\begin{gather}
      \left\{
       \begin{aligned}
        \frac{\partial J_{\alpha}(x)}  {\partial x}
          &=
             - e  \frac{\Delta n_{\alpha}}
                       {\tau_{\alpha\rightarrow\gamma}}
             + e \frac{\Delta n_{\gamma}} {\tau_{\gamma\rightarrow\alpha}}
                 \\
        \frac{\partial J_{\gamma}(x)}  {\partial x}
          &=  -\frac{\partial J_{\alpha}(x)}  {\partial x}
       \end{aligned}
      \right. \quad \Rightarrow
\end{gather}

\begin{gather}
      \left\{
       \begin{aligned}
         \frac{\partial J_{\alpha}(x)}  {\partial x}
          &=
             - e  \frac{N_{\alpha}(E_F)N_{\gamma}(E_F)}
                       {fN_{\alpha}(E_F)+gN_{\gamma}(E_F)}
              \left(
                \frac{g} {\tau_{\alpha\rightarrow\gamma}}
               +\frac{f} {\tau_{\gamma\rightarrow\alpha}}
              \right)
              \left(\mu_{ch,\alpha} - \mu_{ch,\gamma}\right)
                  \\
        \frac{\partial J_{\gamma}(x)}  {\partial x}
          &=  -\frac{\partial J_{\alpha}(x)}  {\partial x}
       \end{aligned}
      \right.
\end{gather}

The above equations can be rewritten in the following form  :

\begin{gather}
      \left\{
       \begin{aligned}
        \frac{\partial J_{\alpha}(x)}  {\partial x}
             &=-L  \left(\mu_{ch,\alpha} - \mu_{ch,\gamma}\right)
              =-L  \left(\mu_{ \alpha} - \mu_{ \gamma}\right)\\
        \frac{\partial J_{\gamma}(x)}  {\partial x}
             &=+L  \left(\mu_{ch,\alpha} - \mu_{ch,\gamma}\right)
              =+L  \left(\mu_{ \alpha} - \mu_{ \gamma}\right)\\
       \end{aligned}
      \right.
      \label{OnsJE2}
\end{gather}

where the coefficient Onsager transport coefficient $L$ is related to
the electronic relaxation times by the following relation:

\begin{center}
      \boxed
      {
       L=e \frac{N_{\alpha}(E_F)N_{\gamma}(E_F)}
                       {fN_{\alpha}(E_F)+gN_{\gamma}(E_F)}
              \left(
                \frac{g} {\tau_{\alpha\rightarrow\gamma}}
               +\frac{f} {\tau_{\gamma\rightarrow\alpha}}
              \right)
      }
      \label{Lmicro}
\end{center}

The form Eq. (\ref{OnsJE2}) is that of Eq. (\ref{con0}) deduced from
the thermokinetic approach in Sec. I, where the coefficient $L$ is
the Onsager coefficient defined in the third equation of Eqs.
(\ref{Onsager0}).

\subsection{Determination of $f$ and $g$}

       We have, for the stationary regime:

       \begin{equation}
          \frac{\partial ( J_{\alpha}(x)-J_{\gamma}(x))}{\partial x}
            =-2L(\mu_{\alpha}-\mu_{\gamma})
       \end{equation}

       Furthermore, the local Ohm's law applied to each channel leads to the
       following equations:

        \begin{gather}
         \left\{
           \begin{aligned}
             J_{\alpha}&=-\frac{\sigma_{\alpha}}{e}
                           \frac{\partial \mu_{\alpha}(x)}{\partial x}\\
             J_{\gamma}&=-\frac{\sigma_{\gamma}}{e}
                           \frac{\partial \mu_{\gamma}(x)}{\partial x}\\
           \end{aligned}
         \right.
         \Rightarrow
          \left\{
           \begin{aligned}
             \frac{\partial J_{\alpha}(x)}{\partial x}
                 &=-\frac{\sigma_{\alpha}}{e}
                     \frac{\partial^2 \mu_{\alpha}(x)}{\partial x^2}\\
             \frac{\partial J_{\gamma}(x)}{\partial x}
                 &=-\frac{\sigma_{\gamma}}{e}
                           \frac{\partial^2 \mu_{\gamma}(x)}{\partial x^2}
          \end{aligned}
         \right.
         \label{Ohm}
        \end{gather}
        where we assume that the conductivities are constant.

       From (\ref{OnsJE2}) and (\ref{Ohm}), we get:
        \begin{gather}
         \left\{
          \begin{aligned}
            \frac{\partial^2 \mu_{\alpha}(x)}{\partial x^2}
              &=\frac{eL}{\sigma_{\alpha}}
                 (\mu_{\alpha}-\mu_{\gamma})\\
            \frac{\partial^2 \mu_{\gamma}(x)}{\partial x^2}
              &=-\frac{eL}{\sigma_{\gamma}}(\mu_{\alpha}-\mu_{\gamma})
          \end{aligned}
         \right.
        \Rightarrow
         \frac{\partial^2  (\mu_{\alpha} -\mu_{\gamma} ) }
              {\partial x^2}
          = eL
            \left(
              \frac{1}{\sigma_{\alpha}}+
              \frac{1}{\sigma_{\gamma}}
            \right)
            (\mu_{\alpha} -\mu_{\gamma} )
          \label{cann}
        \end{gather}
       that leads to the well-known diffusion equation of the chemical
       potential, that describes the {\it spin-accumulation} process:

      \begin{equation}
       \frac{\partial^2  \Delta \mu }
              {\partial x^2}
          = eL
            \left(
              \frac{1}{\sigma_{\alpha}}+
              \frac{1}{\sigma_{\gamma}}
            \right)
            \Delta \mu
      \end{equation}
      from which the spin-diffusion length $l_{sf}$ is deduced:

      \begin{equation}
       \frac{1}{l_{sf}^2}=eL
            \left(
              \frac{1}{\sigma_{\alpha}}+
              \frac{1}{\sigma_{\gamma}}
            \right)
       \label{lsf}
      \end{equation}

      From Eq. (\ref{cann}) we have the differential equations:

      \begin{gather}
        \left\{
          \begin{aligned}
            \frac{\partial^2 \mu_{\alpha}(x)}{\partial x^2}
              &=\frac{eL}{\sigma_{\alpha}}
                 \Delta \mu
                =
                 \frac{\sigma_{\gamma}}{\sigma_t}
                 \frac{\partial^2  \Delta \mu }
                      {\partial x^2}
                 \\
            \frac{\partial^2 \mu_{\gamma}(x)}{\partial x^2}
              &=-\frac{eL}{\sigma_{\gamma}}
                 \Delta \mu
               = -\frac{\sigma_{\alpha}}{\sigma_t}
                  \frac{\partial^2  \Delta \mu }
                       {\partial x^2}
          \end{aligned}
         \right.
      \end{gather}

\subsection{Charge distribution and screening}

      Separating the electric contribution from the chemical contribution,
      the electrochemical potential writes :
       \begin{equation}
         \mu_{\alpha,\gamma}=\mu_{ch,\alpha,\gamma}+eV
       \end{equation}

       so that:
       \begin{gather}
        \left\{
          \begin{aligned}
            \frac{\partial^2 \mu_{\alpha}(x)}{\partial x^2}
              &=  \frac{\partial^2 \mu_{ch,\alpha}(x)}{\partial x^2}
                + e\frac{\partial^2 V(x)}{\partial x^2}\\
            \frac{\partial^2 \mu_{\gamma}(x)}{\partial x^2}
              &=  \frac{\partial^2 \mu_{ch,\gamma}(x)}{\partial x^2}
                + e\frac{\partial^2 V(x)}{\partial x^2}\\
          \end{aligned}
         \right.
        \Rightarrow
        \left\{
          \begin{aligned}
            \frac{\partial^2 \mu_{ch,\alpha}(x)}{\partial x^2}
              &= \frac{\sigma_{\gamma}}{\sigma_t}
                 \frac{\partial^2  \Delta \mu }
                      {\partial x^2}
                + e^2\frac{\Delta n_{\alpha}+\Delta n_{\gamma}}{\epsilon}\\
            \frac{\partial^2 \mu_{ch,\gamma}(x)}{\partial x^2}
              &=-\frac{\sigma_{\alpha}}{\sigma_t}
                  \frac{\partial^2  \Delta \mu }
                       {\partial x^2}
                + e^2\frac{\Delta n_{\alpha}+\Delta n_{\gamma}}{\epsilon}
          \end{aligned}
         \right.
      \end{gather}

      where the Poisson equation has been introduced

      \begin{equation}
        -\frac{\partial^2 V(x) }{\partial x^2}
              =e\frac{\Delta n_{\alpha}+\Delta n_{\gamma}}{\epsilon}
      \end{equation}

The equations rewrite:

\begin{gather}
        \left\{
          \begin{aligned}
            \frac{\partial^2 \delta \mu_{ch,\alpha}(x)}{\partial x^2}
              - e^2\frac{  N_{\alpha} \delta \mu_{ch,\alpha}
                         + N_{\gamma} \delta \mu_{ch,\gamma}
                        }
                           {\epsilon}
              &= \frac{\sigma_{\gamma}}{\sigma_t}
                 \frac{ \Delta \mu }
                      {l_{sf}^2}
                \\
            \frac{\partial^2 \delta \mu_{ch,\gamma}(x)}{\partial x^2}
               -  e^2 \frac{   N_{\alpha} \delta \mu_{ch,\alpha}
                             + N_{\gamma} \delta \mu_{ch,\gamma}
                           }
                           {\epsilon}
              &=-\frac{\sigma_{\alpha}}{\sigma_t}
                  \frac{ \Delta \mu }
                       {l_{sf}^2}
          \end{aligned}
         \right.
      \end{gather}
      These relations with
      \begin{equation}
        \mu_{\alpha}-\mu_{\gamma}=\mu_{ch,\alpha}-\mu_{ch,\gamma}
           =\delta \mu_{ch,\alpha}-\delta \mu_{ch,\gamma}
      \end{equation}
lead to

\begin{gather}
        \left\{
          \begin{aligned}
            \frac{\partial^2 \Delta \mu_{ch,\alpha}(x)}{\partial x^2}
              - \frac{ \Delta \mu_{ch,\alpha}}{l^{2}}
               &= \Delta \mu
                 \left(
                   \frac{\sigma_{\gamma}}{\sigma_t l_{sf}^2}
                 - e^2 \frac{ N_{\gamma}}
                      {\epsilon}
                 \right)
                \\
            \frac{\partial^2 \Delta \mu_{ch,\gamma}(x)}{\partial x^2}
               -  \frac{\Delta \mu_{ch,\gamma}}{l^{2}}
              &=\Delta \mu
                 \left(
                   -\frac{\sigma_{\alpha}}{\sigma_t l_{sf}^2}
                   + e^2 \frac{ N_{\alpha}}
                              {\epsilon}
                 \right)
          \end{aligned}
         \right.
      \end{gather}

      where we have introduced the screening length :

\begin{equation}
         \frac{1}{l^2}=e^2 \frac{N_{\alpha} + N_{\gamma}}{\epsilon}
      \end{equation}

The solution of the equation is composed by a solution of the equation
with zero right hand side (homogeneous solution) and a particular
solution.

\begin{enumerate}
       \item{} \textbf{Solution for \textbf{$\Delta\mu_{ch,\alpha}$}}

        \textit{Homogeneous solution}
         \begin{equation}
          \Delta\mu_{ch,\alpha}^{hmg}= A \, exp \left(\frac{x}{l}\right)
                                     + B \, exp\left(-\frac{x}{l}\right)
         \end{equation}
       \textit{Particular solution}
       $\Delta\mu_{ch,\alpha}^{part}=\mathtt{p_{\alpha}} \Delta \mu$
         \begin{gather}
           \Rightarrow
            \mathtt{p_{\alpha}} \frac{\Delta \mu}{l_{sf}^2}
          -  \frac{\Delta \mu}{l^{2}}
          =
            \left(
              \frac{\sigma_{\gamma}}{\sigma_t l_{sf}^2}
                 - e^2 \frac{ N_{\gamma}}
                      {\epsilon}
            \right)\Delta \mu \\
            \Rightarrow
            \mathtt{p_{\alpha}}=\frac{  \frac {\sigma_{\gamma}}
{\sigma_t l_{sf}^2}
                                      - \frac { e^2 N_{\gamma}} {\epsilon}
                                     }
                                     {  \frac {1} { l_{sf}^2}
                                      -  \frac {1} {l^{2}}
                                     }
         \end{gather}

       \item{} \textbf{Solution for \textbf{$\Delta\mu_{ch,\gamma}$}}

        \textit{Homogeneous solution}
         \begin{equation}
          \Delta\mu_{ch,\gamma}^{hmg}= A' \, exp \left(\frac{x}{l}\right)
                                     + B'\, exp\left(-\frac{x}{l}\right)
         \end{equation}
       \textit{Particular solution:}
$\Delta\mu_{ch,\gamma}^{part}=\mathtt{p_{\gamma}} \Delta \mu$
         \begin{gather}
           \Rightarrow
            \mathtt{p_{\gamma}} \frac{\Delta \mu}{l_{sf}^2}
            - \frac{\Delta \mu}{l^{2}}
          =
            \left(
              \frac{-\sigma_{\alpha}}{\sigma_t l_{sf}^2}
                + e^2 \frac{ N_{\alpha}}
                      {\epsilon}
            \right)\Delta \mu \\
            \Rightarrow
            \mathtt{p_{\gamma}}=\frac{ - \frac {\sigma_{\alpha}}
{\sigma_t l_{sf}^2}
                                       + \frac { e^2 N_{\alpha}} {\epsilon}
                                     }
                                     {  \frac {1} { l_{sf}^2}
                                      -  \frac {1} {l^{2}}
                                     }
         \end{gather}
      \end{enumerate}
      The general solutions satisfying the condition
      \begin{equation}
       \Delta \mu_{ch,\alpha}-\Delta \mu_{ch,\gamma}=\Delta \mu
      \end{equation}
      correspond to
       \begin{equation}
         A=A' \qquad B=B'
       \end{equation}

Inserted in the expression of the charge conservation

\begin{equation}
       f\Delta n_{\alpha}+g\Delta n_{\gamma}=0
\end{equation}

we have

\begin{gather}
      fN_{\alpha}
        \left(
          Aexp\left(\frac{x}{l}\right)+Bexp\left(-\frac{x}{l}\right)
        +\mathtt{p_{\alpha}}\Delta \mu
        \right)
      +gN_{\gamma}
        \left(
          Aexp\left(\frac{x}{l}\right)+Bexp\left(-\frac{x}{l}\right)
        + \mathtt{p_{\gamma}}\Delta \mu
        \right)
      =0 \\ \Rightarrow
      (fN_{\alpha}+gN_{\gamma})
       \left(
          Aexp\left(\frac{x}{l}\right)+Bexp\left(-\frac{x}{l}\right)
        \right)
      +(  fN_{\alpha}\mathtt{p_{\alpha}}
         + gN_{\gamma}\mathtt{p_{\gamma}}
        )
       \Delta \mu
      =0,\ \mathrm{for\ all\ } x \Rightarrow
\end{gather}

\begin{equation}
      fN_{\alpha}\mathtt{p_{\alpha}}
              + gN_{\gamma}\mathtt{p_{\gamma}}
              = 0
             \qquad \mathrm{and}\qquad
             A=B=0
\end{equation}
\begin{equation}
       \Rightarrow
       fN_{\alpha} \left(  \frac {\sigma_{\gamma}} {\sigma_t l_{sf}^2}
                         - \frac { e^2 N_{\gamma}} {\epsilon}
                   \right)
      + gN_{\gamma} \left(  -  \frac {\sigma_{\alpha}} {\sigma_t l_{sf}^2}
                            +  \frac { e^2 N_{\alpha}} {\epsilon}
                    \right)
      =0
\end{equation}
A solution writes:

\begin{gather}
        \left\{
          \begin{aligned}
            f&=N_{\gamma} \left(    \frac {\sigma_{\alpha}} {\sigma_t l_{sf}^2}
                                -  \frac { e^2 N_{\alpha}} {\epsilon}
                         \right)\\
            g&=N_{\alpha} \left(  \frac {\sigma_{\gamma}} {\sigma_t l_{sf}^2}
                               - \frac { e^2 N_{\gamma}} {\epsilon}
                         \right)
          \end{aligned}
         \right.
      \end{gather}

\subsection{$l_{sf}$ function of $l_{\alpha}$,$l_{\gamma}$ and $l$}

      The relation between $L$ and the electronic relaxation times has
      been found to be:

      \begin{equation}
        L=e \frac{N_{\alpha}(E_F)N_{\gamma}(E_F)}
                       {fN_{\alpha}(E_F)+gN_{\gamma}(E_F)}
              \left(
                \frac{g} {\tau_{\alpha\rightarrow\gamma}}
               +\frac{f} {\tau_{\gamma\rightarrow\alpha}}
              \right)
      \end{equation}

       Inserting the expression of $f$ and $g$ obtained in the previous
       paragraph, $L$ becomes:
       \begin{equation}
        L=e \frac{  \frac{1}{\sigma_t l^2_{sf}}
                    \left(  \frac{\sigma_{\alpha} N_{\gamma}}
                                 {\tau_{\gamma\rightarrow\alpha}}
                          + \frac{\sigma_{\gamma} N_{\alpha}}
                                 {\tau_{\alpha\rightarrow\gamma}}
                    \right)
                  - e^2\frac{ N_{\alpha}N_{\gamma} } {\epsilon}
                    \left(
                       \frac{1}{\tau_{\alpha\rightarrow\gamma}}
                     + \frac{1}{\tau_{\gamma\rightarrow\alpha}}
                    \right)
                 }
                 {  \frac{1}{l_{sf}^2}
                  - \frac{1}{l^2 }
                 }
       \end{equation}

      Furthermore, according to Eq.  (\ref{lsf}), the coefficient $L$
      can also be written in the following form:
      \begin{equation}
        L=\frac{\sigma_{\alpha}\sigma_{\gamma}}
               {e\, \sigma_{t}\, l_{sf}^2}
      \end{equation}

      Both results lead to the equation:

      \begin{equation}
        e^2 \frac{  \frac{1}{\sigma_t l^2_{sf}}
                    \left(  \frac{\sigma_{\alpha} N_{\gamma}}
                                 {\tau_{\gamma\rightarrow\alpha}}
                          + \frac{\sigma_{\gamma} N_{\alpha}}
                                 {\tau_{\alpha\rightarrow\gamma}}
                    \right)
                  - e^2\frac{ N_{\alpha}N_{\gamma} } {\epsilon}
                    \left(
                       \frac{1}{\tau_{\alpha\rightarrow\gamma}}
                     + \frac{1}{\tau_{\gamma\rightarrow\alpha}}
                    \right)
                 }
                 {  \frac{1}{l_{sf}^2}
                  - \frac{1}{l^2 }
                 }
        =
        \frac{\sigma_{\alpha}\sigma_{\gamma}}
               {\, \sigma_{t}\, l_{sf}^2}
          \label{difflength}
      \end{equation}

      Let us define the typical diffusion length per channel $l_{\alpha}$
      et $l_{\gamma}$ such that
      \begin{gather}
       \left\{
        \begin{aligned}
          l_{\alpha}^2&=\frac{\sigma_{\alpha}} {e^2N_{\alpha}}
                       \tau_{\alpha\rightarrow\gamma}\\
          l_{\gamma}^2&=\frac{\sigma_{\gamma}} {e^2N_{\gamma}}
                       \tau_{\gamma\rightarrow\alpha}\\
        \end{aligned}
       \right.
      \Rightarrow
      \left\{
        \begin{aligned}
          \frac{1}{\tau_{\alpha\rightarrow\gamma}}
             &=\frac{\sigma_{\alpha}} {e^2N_{\alpha} l_{\alpha}^2}\\
          \frac{1}{\tau_{\gamma\rightarrow\alpha}}
             &=\frac{\sigma_{\gamma}} {e^2N_{\gamma} l_{\gamma}^2}
        \end{aligned}
      \right.
      \Rightarrow
      \left\{
        \begin{aligned}
          \frac{\sigma_{\alpha} N_{\gamma}} {\tau_{\gamma\rightarrow\alpha}}
             &=\frac{\sigma_{\gamma} \sigma_{\alpha}} {e^2  l_{\gamma}^2}  \\
          \frac{\sigma_{\gamma} N_{\alpha}} {\tau_{\alpha\rightarrow\gamma}}
            &=\frac{\sigma_{\gamma} \sigma_{\alpha}} {e^2  l_{\alpha}^2}
        \end{aligned}
       \right.
      \end{gather}

      Eq. (\ref{difflength}) rewrites:
      \begin{gather}
        \frac{ \frac{\sigma_{\alpha} \sigma_{\gamma}}
                    {e^2 \sigma_t l^2_{sf}}
                    \left(  \frac{1} {l_{\gamma}^2}
                          + \frac{1}{l_{\alpha}^2}
                    \right)
                  -  \frac{ N_{\alpha}N_{\gamma} } {\epsilon}
                    \left(
                       \frac{\sigma_{\alpha}} { N_{\alpha} l_{\alpha}^2}
                     + \frac{\sigma_{\gamma}} { N_{\gamma} l_{\gamma}^2}
                    \right)
                 }
                 {  \frac{1}{l_{sf}^2}
                  - \frac{1}{l^2 }
                 }
        =
        \frac{\sigma_{\alpha}\sigma_{\gamma}}
               { e^2 \, \sigma_{t}\, l_{sf}^2}
        \Rightarrow\\
        \frac{ \sigma_{t}\,} {N_{\alpha}+N_{\gamma}}
             \left(
               \frac{N_{\gamma}} { \sigma_{\gamma} l_{\alpha}^2}
                     + \frac{N_{\alpha}} { \sigma_{\alpha} l_{\gamma}^2}
             \right) l_{sf}^4
        -\left[
           l^2 \left(
                 \frac{1} {l_{\gamma}^2} +  \frac{1}{l_{\alpha}^2}
               \right)
            +1
         \right] l_{sf}^2
       + l^2
       =0
      \end{gather}
\paragraph{Limits:}

In metals, the screening length is much smaller than the
diffusion length of both channels:

\begin{enumerate}
       \item{$l/l_{\alpha \gamma} \ll 0$}
         \begin{equation}
          \frac{ \sigma_{t}\,} {N_{\alpha}+N_{\gamma}}
             \left(
               \frac{N_{\gamma}} { \sigma_{\gamma} l_{\alpha}^2}
                     + \frac{N_{\alpha}} { \sigma_{\alpha} l_{\gamma}^2}
             \right) l_{sf}^2
          -1=0 \Rightarrow
          \boxed
          { \frac{1}{l_{sf}^2}
               = \frac{N_{\gamma}/(N_{\alpha}+N_{\gamma})}
                     {\sigma_{\gamma}/(\sigma_{\alpha}+\sigma_{\gamma})}
                 \frac{1}{l_{\alpha}^2}
               + \frac{N_{\alpha}/(N_{\alpha}+N_{\gamma})}
                     {\sigma_{\alpha}/(\sigma_{\alpha}+\sigma_{\gamma})}
                 \frac{1}{l_{\gamma}^2}
          }
        \end{equation}

        The other limit gives:
       \item{$ l/l_{\alpha \gamma} \Rightarrow \infty$}
         \begin{equation}
          l^2 \left[\
                1-\left(
                     \frac{1}{l_{\alpha}^2} +  \frac{1} {l_{\gamma}^2}
                  \right)
                  l_{sf}^2
              \right]
              =0 \Rightarrow
         \frac{1}{l_{sf}^2}=\frac{1}{l_{\alpha}^2} +  \frac{1} {l_{\gamma}^2}
        \end{equation}
\end{enumerate}

But this second limit is not expected in usual materials.

\section{Onsager matrix}
\addtocounter{equation}{141}
   \markboth{\small{A. ONSAGER MATRIX}}{}
         The aim of this Appendix is to derive the Onsager matrix
         (\ref{OnsagerF0}), (and as a particular case (\ref{Onsager0})) on
         the basis of the first and second laws of thermodynamics.  In a
         typical one dimensional junction the layer is decomposed into
         $\Omega$ parts, defining the sub-system $ \Sigma^k $, which is in
         contact to the ``reservoirs'' $ \Sigma^{k-1} $ and $ \Sigma^{k+1}
$.
         The sub-systems $ \Sigma^k $, is then an open system which
exchanges
         heat and chemical species with its left and right vicinity layers.
         Furthermore, the populations ($N^k_{s \uparrow}$, $N^k_{s
         \downarrow}$, $N^k_{d \uparrow}$) and spin down ($N^k_{d
         \downarrow}$) are not conserved due to transitions from one channel
         to the other.

In this picture, the states of the sub-system $ \Sigma^k $
        are described by the variables

\begin{equation}
(S^k, N^k_{s \uparrow},N^k_{s \downarrow}, N^k_{d \uparrow},N^k_{d
\downarrow}) \label{var}
\end{equation}
where $S^k$ is the entropy. The internal variables $\Psi_{s}$,
$\Psi_{d}$ and $\Psi_{sd}$ must however be introduced in order to
take into account the relaxation processes occurring respectively
between the two s-spin channels, the two d-spin channels, and the
s-d relaxation .

Let us define the heat and chemical power by $P_{\phi}$ (the
mechanical power is zero as long as the action of the magnetic field
on the charge carriers is neglected).  The {\it first law of the
thermodynamics} applied to the layer $\Sigma^k$ gives

\begin{equation}
\frac{dE^k}{dt}\,=\,P_{\phi}^{k-1 \to k}\,-\,P_{\phi}^{k \to k+1}
\end{equation}

Introducing the canonical definitions $T^k=\frac{\partial
E^k}{\partial S^k} $ and  $ \mu_{s\pm}^k=\frac{\partial
E^k}{\partial N_{s\pm}^k}, \, \mu_{d\pm}^k=\frac{\partial
E^k}{\partial N_{d\pm}^k}\,$ the energy variation is:

\begin{equation}
\frac{dE^k}{dt}\,=\,T^k \frac{dS^k}{dt}\,+\, \mu_{s \uparrow}^k
\frac{dN_{s\uparrow}^k}{dt}\, +\, \mu_{s \downarrow}^k \frac{dN_{s
\downarrow}^k}{dt} \, + \, \mu_{d \uparrow}^k \frac{dN_{d
\uparrow}^k}{dt}\, +\, \mu_{d \downarrow}^k \frac{dN_{d
\downarrow}^k}{dt} \label{firstPrin}
\end{equation}

For the sake of simplicity,
        we limit our analysis to the isothermal case,
$T^k=T$.  The entropy variation of the sub-layer is deduced from the
two last equations, after introducing the conservation laws and
after defining the polarized currents  $\delta I_{\downarrow } \, =
\, I_{s \downarrow } - I_{d \downarrow} $, $\delta
I_{\downarrow} \, = \, I_{s \downarrow } - I_{d \downarrow} $,
and the currents $I_{\downarrow} \, =  I_{s \downarrow } + I_{d
\downarrow} $,  $I_{s} \, = I_{s \uparrow } + I_{s
\downarrow}$,

\begin{eqnarray}
T\frac{dS^k}{dt}\,& = &\, \,P_{\phi}^{k-1 \to k}\,-\,P_{\phi}^{k \to
k+1} \, -\, \frac{1}{2} \Delta \mu_{s} ^k \left(\delta I^{k-1 \to
k}_{\downarrow}- \delta I^{k \to k+1}_{\downarrow} \,+\,
\dot{\Psi}_{sd}^k - 2 \, \dot{\Psi}_{s}^k \right)
\, \nonumber\\
& & -\, \frac{1}{2} \mu^k_{s} \left (I^{k-1 \to k}_{s}-I^{k \to
k+1}_{s} - \dot{\Psi}_{sd}^k \right ) -\, \frac{1}{2} \Delta
\mu_{\downarrow} ^k \left( \delta I^{k-1 \to k}_{\downarrow}- \delta
I^{k \to k+1}_{\downarrow}  - 2 \dot{\Psi}_{sd}^k - \dot{\Psi}_{s}^k
\right)
\,  \nonumber\\
& & -\, \frac{1}{2} \mu^k_{\downarrow} \left ( I^{k-1 \to
k}_{\downarrow}-I^{k \to k+1}_{\downarrow} + \dot{\Psi}_{s}^k
\right) \label{entropy}
\end{eqnarray}
where we have introduce the chemical potentials $\mu_{s}^k\,\equiv
\,\mu_{s \uparrow}^k\, +\, \mu_{s \downarrow}^k/2$,
$\mu_{\downarrow}^k\,\equiv \,\mu_{s \downarrow}^k/2\, +\, \mu_{d
\downarrow}^k$, and {\it the chemical affinities of the reactions},
defined by $ \Delta \mu_{s}^k\,\equiv \, \mu_{s \uparrow}^k-\mu_{s
\downarrow}^k/2 =- \frac{\partial E^k}{\partial \Psi_{s}^k}$, $ \Delta
\mu_{\downarrow}^k\,\equiv \, \mu_{s \downarrow}^k/2-\mu_{d
\downarrow}^k = -\frac{\partial E^k}{\partial \Psi_{sd}^k}$.

The entropy being an extensive variable, the total entropy variation
of
        the system is obtained by summation over the layers 1 to $\Omega$
        where the layer 1 is in contact to
the left reservoir $R^l$ and the layer $\Omega$ is in contact to the
        right reservoir $R^r$.

The total entropy variation is:

\begin{eqnarray}
T\frac{dS}{dt}\,& = &\, [P]^{R^l \to 1} - [P]^{\Omega \to
R^r} \nonumber\\
& & +\, \sum_{k=2}^{\Omega}\frac{1}{2} \left(\Delta
\mu_{s}^{k-1}-\Delta \mu_{s}^k \right ) \, \delta I^{k-1 \to k}_{s}
+\, \sum_{k=2}^{\Omega} \frac{1}{2} (\mu_{s}^{k-1} - \,
\mu_{s}^{k})\, I^{k-1 \to k}_{0 s} \nonumber\\
& & + \, \sum_{k=2}^{\Omega}\frac{1}{2} \left(\Delta
\mu_{\downarrow}^{k-1}-\Delta \mu_{\downarrow}^k \right ) \, \delta
I^{k-1 \to k}_{\downarrow}
        +  \nonumber\\
& & \sum_{k=2}^{\Omega} \frac{1}{2} (\mu_{\downarrow}^{k-1} \,- \,
\mu_{\downarrow}^{k})\, I^{k-1 \to k}_{0 \downarrow} \, + \,
\sum_{k=1}^{\Omega} \Delta \mu_{s}^{k} \, \dot{\Psi}_{s}^k \,+
\sum_{k=1}^{\Omega} \Delta \mu_{\downarrow}^{k} \, \dot{\Psi}_{sd}^k
\label{entropytot2}
\end{eqnarray}

where $[P]^{R_{l} \to 1}$ and $[P]^{\Omega \to R_{r}} $ in the right
hand side of the equality stand for the heat and chemical transfer
from the reservoirs to the system $ \Sigma $.

The entropy variation takes the form

\begin{equation}
T\frac{dS}{dt}\, = \,\sum_{i}F_{i} \dot{X}^{i} \,+\, P^{ext}(t)
\end{equation}
where $F_{i}$ are generalized forces and $\dot{X}^{i} $ are the
conjugate generalized fluxes. The variation of entropy is composed
of an external entropy variation $P^{ext}(t)/T$ and by an internal
entropy variation $\mathcal I$.

By applying {\it the second law of thermodynamics} $\mathcal I \ge
0$ we
        are introducing the kinetic
coefficients $L_{i j }$ such that $\mathcal I=\sum_{
i}F_{i} \left( \sum_{ j} L_{ij} F^{j} \right)$. By identification
with the expression~(\ref{entropytot2}), the kinetic equations are
obtained, after performing the continuous limit,

\begin{eqnarray}
\left[\begin{array}{c}
J_{0s} \\
J_{0 \downarrow} \\
\delta J^d_{s}\\
\delta J^d_{\downarrow}\\
\dot{\psi}_{s} \\
\dot{\psi}_{\downarrow} \\
\end{array}\right]
= \left[\begin{array}{cccccc}
L_{ss} & L_{s\downarrow}  & 0  & 0 & 0   & 0\\
L_{\downarrow s} & L_{\downarrow \downarrow}  & 0   & 0  & 0 & 0\\
0  & 0  & L_{\delta s \delta s} & L_{\delta s \delta \downarrow} & 0
& 0\\
0  & 0  & L^d_{\delta \downarrow \delta s} & L^d_{\delta \downarrow
\delta \downarrow}  & 0  & 0\\
0  & 0  & 0 & 0  &  L_{int}^{s}   & 0 \\
0  & 0  & 0 & 0  & 0 & L_{int}^{ \downarrow} \\
\end{array}\right]
\left[\begin{array}{c}
        \frac{-\partial \mu_{s}}{\partial z}\\
        \frac{-\partial \Delta \mu_{s}}{\partial z} \\
        \frac{-\partial \mu{\delta }}{\partial z}\\
        \frac{-\partial \Delta \mu_{\downarrow}}{\partial z} \\
        {\Delta \mu_{s}} \\
        {\Delta \mu_{\downarrow}} \\
\end{array}\right]
\label{kineticElec}
\end{eqnarray}

The kinetic coefficients are state functions; $L_{ij} = L_{ij }(S^k,
N^k_{+},N^k_{-})$ and the symmetrized matrix is positive : $
\frac{1}{2}\,\left\{L_{ji}\,+\,L_{ij} \right\}_{\{ij\}} \ge 0$. The
coefficients $L_{int}$ refer to the internal relaxation processes
\cite{DeGroot,Mazur}. According to Onsager reciprocity relations, the kinetic
coefficients are symmetric or antisymmetric $L_{ij}=\pm L_{ji}$. The
coefficients are known from the two-channel model for the
conductivity. The two last equations concern the internal
($L_{int}$) ``density'' variables $\psi_{s} $ and $\psi_{sd} $
defined by $\Psi^k = \int_{\Sigma^k} \, \psi(z)dz $.  Due to the
Curie principle, there is no coupling between spin polarized
transport processes and the electronic transitions (the scalar
process is not coupled to vectorial processes).


\begin{thebibliography}{kfjgdslkjyg}
\bibitem{RquChim} the coefficient $\nu_{ij}$ is defined by the
stochiometric coefficients. If the chemical reaction $i$ writes
$\nu'_{iA} A \rightarrow \nu''_{iA} A$, with stochiometric
coefficients $\nu'_{iA}$ and $\nu''_{iA}$, the coefficent $\nu_{iA}$
apprearing in the continuity equations is $\nu_{iA} =\nu''_{iA} - \nu'_{iA}$.
\bibitem{RqueEcin} In what follows, there is no kinetic energy, i.e. no
inertial effects. This is the reason why the equations are reduced to
the simplest expression.
\bibitem{Constantes} In all these calculations, the channel parameters
$\sigma_{\alpha \gamma}$, $\beta$, and $L$ are considered as
constant in space. It can be shown that taking into account the
gradient only introduces vanishingly small corrections.
\bibitem{Integral} The Riemann integral is evaluated in the intervals
where the derivative is regular. Note that a global calculation over
$\left[ A,B\right] $, within the framework of the distribution
theory, yields zero because $\Delta \mu (A)=\Delta \mu (B)=0$. This
simply means that the integral of the regular part is opposite to
the integral of the Dirac masses located at the interface. Let us
write that the out-of-equilibrium resistance is straightforwardly
connected to the chemical potential drop between the A and B points,
diminished by the standard potential drop which would result from
the application of Ohm's law. We obtain : $J_{t}eR^{ne} = \left[ \mu
_{I}(A)-\mu _{II}(B)\right] -e\left[ \Phi _{I}(A)-\Phi
_{I}(0^{-})\right] -e\left[ \Phi _{II}(0^{+})-\Phi _{II}(B)\right]
=\left[ \mu _{I}(A)-e\Phi _{I}(A)\right] -\left[ \mu (0)-e\Phi
_{I}(0^{-}) \right] +\left[ \mu (0)-e\Phi _{II}(0^{+})\right]
-\left[ \mu _{II}(B)-e\Phi _{II}(B)\right] $ $=e\left[ \Phi
_{I}(0^{-})-\Phi _{II}(0^{+})\right] $. It can be seen that the
second expression for $J_{t}eR^{ne}$ above is nothing but the
opposite of Eq.(6), and is equal to the discontinuity of $\Phi $ at
the interface, which provides a simple physical interpretation as
illustrated in Fig. \ref{chimique}.
\bibitem{DmuVal} According to Eq.  (\ref{Result}), the value of
$\Delta \mu$ is of the order of the giant magnetoresistance
multiplied by the current, $\Delta R. I$, i.e. of the order of tens
of $\mu eV$ (up to $meV$ in CIMS experiments).
\end{thebibliography}
\end{document}